%% file: iclr2027_conference.tex
\documentclass{article} 
\usepackage{iclr2027_conference,times}

\usepackage[utf8]{inputenc}
\usepackage[T1]{fontenc}
\usepackage{hyperref}
\usepackage{url}
\usepackage{booktabs}
\usepackage{amsfonts}
\usepackage{amsmath}
\usepackage{nicefrac}
\usepackage{microtype}
\usepackage{xcolor}
\usepackage{enumitem}
\usepackage{algorithm}
\usepackage{algpseudocode}
\usepackage{tikz}
\usetikzlibrary{positioning, arrows.meta, decorations.pathreplacing, calc}
\usepackage{graphicx}
\usepackage{float} 

\usepackage{fvextra}
\usepackage[most]{tcolorbox}
\usepackage[framemethod=default]{mdframed}
\usepackage{needspace}
\usepackage{soul}
\usepackage{subcaption}
\usepackage{placeins}
\DefineVerbatimEnvironment{promptverbatim}{Verbatim}{
  breaklines=true,
  fontsize=\small,
  breakanywhere=true,
  breaksymbolleft=,
  breaksymbolright=,
}

\definecolor{diffamber}{HTML}{F1D27A}
\definecolor{diffgreen}{HTML}{B8DDB5}
\definecolor{diffred}{HTML}{C62828}
\definecolor{utilityprofile}{HTML}{F8766D}
\definecolor{utilitycode}{HTML}{7CAE00}
\definecolor{utilitygrid}{HTML}{00BFC4}
\newcommand{\diffreplace}[1]{{\sethlcolor{diffamber}\hl{#1}}}

\newcommand{\diffdelete}[1]{{\setstcolor{diffred}\textcolor{diffred}{\st{#1}}}}
\DeclareRobustCommand{\dashedunderline}[1]{%
  \tikz[baseline=(text.base)]{%
    \node[inner sep=0pt,outer sep=0pt] (text) {#1};%
    \draw[dashed,line width=0.45pt] ([yshift=-1.1pt]text.south west) -- ([yshift=-1.1pt]text.south east);%
  }%
}

\mdfdefinestyle{prompttemplatestyle}{
  backgroundcolor=gray!4,
  linecolor=gray!55!black,
  linewidth=0.45pt,
  innerleftmargin=4pt,
  innerrightmargin=4pt,
  innertopmargin=3pt,
  innerbottommargin=3pt,
  skipabove=5pt,
  skipbelow=7pt,
  splittopskip=\baselineskip,
  splitbottomskip=2pt,
  frametitlebackgroundcolor=gray!12,
  frametitlerule=true,
  frametitlerulecolor=gray!55!black,
  frametitlerulewidth=0.45pt,
  frametitlefont=\footnotesize\sffamily\bfseries,
  frametitleaboveskip=2pt,
  frametitlebelowskip=2pt,
}

\newenvironment{prompttemplate}[1]{%
  \begin{mdframed}[style=prompttemplatestyle,frametitle={#1},nobreak=true]%
}{%
  \end{mdframed}%
}

\title{LLM Anonymization Against Agentic Re-Identification}

\iclrfinalcopy

\author{%
  Ziwen Li \\
  Khoury College of Computer Sciences\\
  Northeastern University\\
  Boston, MA \\
  \texttt{li.ziw@northeastern.edu} \\
  \And
  Jianing Wen \\
  Khoury College of Computer Sciences\\
  Northeastern University\\
  Boston, MA \\
  \texttt{wen.jiani@northeastern.edu} \\
  \And
  Tianshi Li \\
  Khoury College of Computer Sciences\\
  Northeastern University\\
  Boston, MA \\
  \texttt{tia.li@northeastern.edu} \\
}

\begin{document}

\maketitle

\begin{abstract}
Agentic LLMs with web search change the threat model for text anonymization: weak contextual cues can become cross-referenceable evidence for re-identification, yet those same details also carry downstream analytic value of the text.
Existing defenses either remove explicit identifiers, perturb text for formal privacy, or test rewritten text against non-web inference models, leaving underexplored the operating region between resistance to agentic web-search re-identification and utility retention.
We introduce AURA (\textbf{A}nonymization with \textbf{U}tility-\textbf{R}etention \textbf{A}daptation), an LLM-powered \textit{mask-reconstruct} framework that decouples privacy localization from utility-preserving reconstruction and selects candidates with adversarial privacy and utility-retention checks.
We evaluate AURA on real-user interview transcripts using re-identification attacks carried out by web-search agents, along with a utility evaluation based on interviewee-profile facts, codebook facts, and the joint contextual utility grid.
Our results show that adaptive-scope AURA yields the lowest agentic re-identification counts under each of three attacker models among the non-DP methods, and that at matched scope and backbone, AURA's mask-reconstruct design retains more contextual utility than the prior LLM anonymizer (+6.4 pp unit-grid recovery) at comparable privacy
\footnote{Source Code: \url{https://github.com/AaronLi43/AURA}}.
\end{abstract}

\input{sections/1_introduction}

\input{sections/2_related}
\input{sections/3_method}

\input{sections/4_experiments}

\input{sections/5_conclusion}
\input{sections/AI_usage}


\bibliographystyle{iclr2027_conference}
\bibliography{iclr2027_conference}

\newpage
\appendix
\input{sections/appendix}

\end{document}

%% file: sections/1_introduction.tex
\section{Introduction}
\label{sec:intro}

The rapid growth in large language models (LLMs)~\citep{achiam2023gpt, yang2025qwen3technicalreport} capabilities and adoption has renewed debate about their societal impact, with privacy emerging as a central concern beyond early work on training-data memorization.
Recent works have shown that modern agentic LLM systems can use web search, generate queries from weak contextual cues, retrieve public evidence, and cross-reference external materials to infer identities~\citep{staab2024beyond,li2026deanonymizers,lermen2026largescaleonlinedeanonymizationllms,ko2026weakcuesrealidentities}.
This makes LLM-era anonymization less a problem of deleting explicit identifiers than of deciding which contextual details can safely remain.


The stakes are amplified by the growing wave of large-scale AI use data collection and re-distribution. Public datasets such as LMSYS-Chat-1M~\citep{zheng2023lmsys}, WildChat~\citep{zhao2024wildchat}, SWE-chat~\citep{baumann2026swechat}, Anthropic Interviewer~\citep{anthropic2025interviewer}, release increasingly rich records of how people use AI systems, while providers conduct proprietary analyses that require the same richness (e.g., Anthropic's Clio~\citep{tamkin2024clio}, the Anthropic Economic Index \citep{appel2025anthropic}, and OpenAI's NBER study~\citep{chatterji2025people}).
These efforts share a common requirement: research questions that cut across respondent background, behavioral patterns, domain expertise, and attitudinal reasoning demand data that retains contextual nuance across multiple dimensions simultaneously. Yet the same contextual details that carry analytic value are precisely the cues that agentic LLMs exploit for re-identification~\citep{li2026deanonymizers, lermen2026largescaleonlinedeanonymizationllms, ko2026weakcuesrealidentities}, creating a structural tension between the privacy of participants and the utility of the provided data.

Current anonymization methods are inadequate for this challenge.
Heuristic sanitization methods represent the status quo in real-world data sanitization practices. In particular, Named-entity recognition (NER) systems such as Microsoft Presidio~\citep{microsoft2021presidio} detect explicit identifiers (names, emails, dates) but miss the contextual inference cues that modern LLMs exploit~\citep{staab2024beyond}.
LLMs demonstrate potential to advance anonymization by hiding sensitive attributes from model inferences~\citep{staab2025anonymizers} beyond explicit identifier redaction.
As for formal privacy guarantee, differential privacy is the predominant framework in machine learning, but DP text rewriting methods~\citep{meisenbacher2024dpmlm, mattern2022limits, utpala2023dpprompt} often achieve these guarantees by token-level perturbation which can substantially degrade readability and analytical value.
Practical DP deployment requires theoretically managing privacy-utility tradeoffs using privacy budget parameters as tuning knobs~\citep{papernot2022hyperparameter, 1074842, NEURIPS2023_59b9582c} and investigating the impact on privacy and utility empirically~\citep{pmlr-v267-hu25x}.

We ask this question: \textit{How can we anonymize text against the pragmatic agentic re-identification threats while preserving the information needed for downstream use?}
It boils down to two gaps.
\textbf{First}, agentic attacks make sensitive span localization \textit{beyond predefined categories} necessary because the attack success is contingent on the availability of contextual cues which not only include typical personal attributes but also reflect niche, idiosyncratic details that only become identifying when combined with external evidence. 
\textbf{Second}, identifying these cues is not enough. Unlike explicit identifiers, contextual cues that create re-identification risk often carry substantive downstream insights.
Anonymization must therefore decide not only which spans are risky but how to transform them to preserve utility.


Thus, we introduce \textbf{AURA} (\textbf{A}nonymization with \textbf{U}tility-\textbf{R}etention \textbf{A}daptation), an LLM-powered \emph{mask-reconstruct} framework for LLM-era text anonymization.
AURA separates privacy localization from utility-preserving reconstruction: it first identifies contextual cues that may support agentic web-search re-identification, then reconstructs the affected spans under empirical constraints and metrics of privacy and utility.
We evaluate AURA and other anonymization methods on real-world interview transcripts from the Anthropic Interviewer dataset~\citep{anthropic2025interviewer} verified as vulnerable to agentic re-identification.
We applied AURA variants under closed-source and open-weight LLM backbones, along with five other anonymization methods to these transcripts and then tested the outputs against agentic re-identification attacks.
Across three deanonymization attacker models (\texttt{GPT-5.4-mini}, \texttt{GPT-5.1}, and \texttt{Gemini-3-Flash}), AURA's adaptive-privacy variants reduce agentic re-identification to 0-7/53 transcripts, substantially below NER-based redaction (26-40/53) and consistently lower across all three attackers than the prior LLM-based anonymizer~\citep{staab2025anonymizers} (10-12/53) while retaining 72.0-76.1\% of unit-level utility-grid information. Open-weight backbones such as \texttt{Qwen3.5-27B} and \texttt{Qwen3.5-35B-A3B}, which allows local deployment, match or exceed the API-powered baseline on utility (76.0\%/76.1\% vs.\ 72.0\% unit-grid recovery) at comparable privacy.

Our main contributions are:
\begin{enumerate}[leftmargin=*,itemsep=2pt,topsep=2pt]
    \item To our knowledge, we are the first to optimize and evaluate LLM text anonymization in the operating region between resistance to real-world agentic web-search re-identification and retention of downstream analytic utility.
    \item We propose \textbf{AURA}, an LLM-powered \textit{mask-reconstruct} framework that decouples \emph{where} to intervene from \emph{how} to rewrite, and is evaluated with both agentic re-identification attacks and utility-retention checks.
    \item We empirically characterize how privacy and utility change across anonymization settings; our results are consistent with scope design primarily improving resistance to re-identification and mask-reconstruct preserving utility at comparable privacy.
\end{enumerate}

\begin{figure}[tb]
    \centering
    \includegraphics[width=0.99\linewidth]{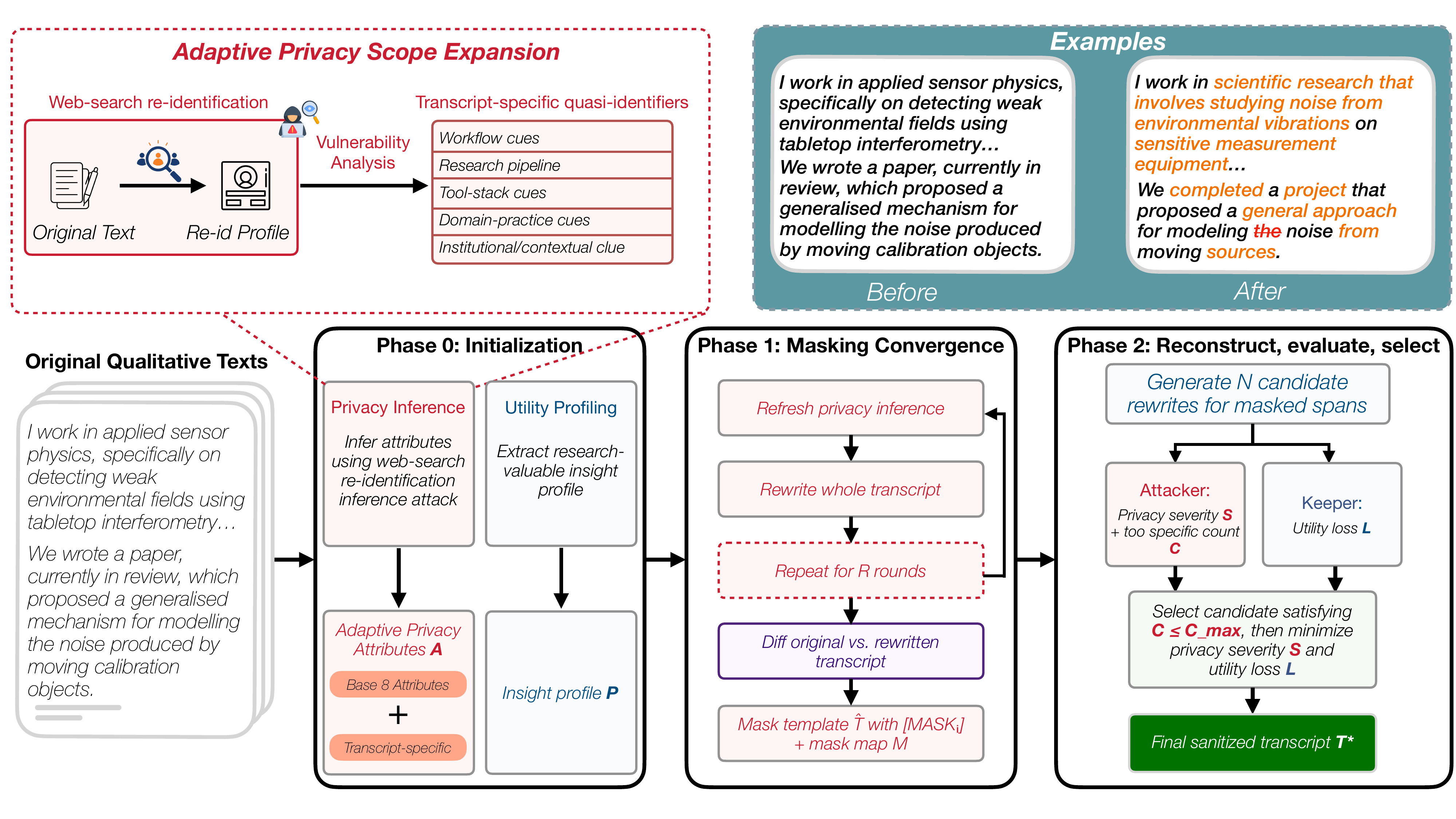}
    \caption{AURA overview. Adaptive privacy scope expansion first augments a base re-identification profile with transcript-specific quasi-identifiers, then AURA initializes privacy and utility profiles, converges on a mask template for sensitive spans, and reconstructs only those spans based on the masked original transcript. Candidate rewrites are evaluated by an attribute inference attacker and a keeper before selecting the final sanitized transcript $T^*$.}
    \label{fig:pipeline}
\end{figure}

%% file: sections/2_related.tex
\section{Related Work}
\label{sec:related}

Text de-identification has long focused on the removal of explicit identifiers based on predefined taxonomies.
The related technical problem is named entity recognition (NER), which has progressed from rule-based systems to neural models that approach human performance on standard benchmarks~\citep{meystre2010deid, stubbs2015i2b2, dernoncourt2017deid, niklaus2023legal}.
Publicly available NER tools such as Presidio~\citep{microsoft2021presidio} have been widely cited in research to support anonymization in textual data release~\citep{zheng2023lmsys, zhao2024wildchat, baumann2026swechat, lin2023toxicchat}.
However, releasing qualitative data containing detailed behavioral or attitudinal textual data requires higher standards for anonymization because interview transcripts contain free-form disclosures whose identifying power emerges from context and attribute combinations~\citep{narayanan2008robust, lison2021anonymisation, pilan2022text}, which named-entity removal alone cannot address.
This type of data is usually treated with high caution and traditionally based on manual anonymization~\citep{saunders2015anonymising, surmiak2018confidentiality, heaton2008secondary, bishop2009ethical}, which remains ad-hoc and impractical to scale, and can especially become vulnerable to intensified real-world re-identification threats that are democratized and scalable by LLMs~\citep{li2026deanonymizers, ko2026weakcuesrealidentities, lermen2026largescaleonlinedeanonymizationllms}.
AURA directly tackles this gap by localizing and transforming contextual quasi-identifiers in long-form transcripts rather than only deleting named entities.

Recent work has begun to use LLMs for text sanitization and anonymization, some explicitly modeling privacy-utility trade-offs~\citep{yang2025robust, siyan-etal-2025-papillon, zhou2025operationalizing, staab2025anonymizers}.
However, these works operationalize privacy and utility in qualitatively different ways.
Some address privacy leakage to remote LLM providers by redacting~\citep{siyan-etal-2025-papillon} or abstracting~\citep{siyan-etal-2025-papillon, zhou2025operationalizing} NER-based PII before prompts are shared.
A second, more nascent line of work studies inferential privacy risks~\citep{mireshghallah2025position}, proposing defenses that include iteratively rewriting text to suppress personal attributes from being inferrable~\citep{staab2025anonymizers} or prevent non-agentic LLM-based re-identification~\citep{yang2025robust}.
Our work targets a distinct and stronger threat model: agentic re-identification, where an LLM agent combines textual cues with web search evidence to identify ordinary individuals with linkable online traces. This setting is both more dynamic and more costly to defend against, making it impractical to place an agentic attacker inside every rewriting iteration.
AURA's mask-reconstruct framework provides a solution by decoupling the problem into two parts: 1) a one-off run of an agentic attacker to identify attributes that can be used to re-identify the interviewee; 2) an iterative rewriting process that only uses LLM-based attribute inference to check the preservation of the attributes resulting from the first step.
To establish a direct comparison, we included \citet{staab2025anonymizers} and two one-shot LLM rewriting settings as baselines.

At the formal end of the design space, differential privacy (DP) rewrites text via synthetic representations, word-level noise, or DP-fine-tuned generators~\citep{dwork2014algorithmic, weggenmann2018syntf, mattern2022limits, igamberdiev2023dpbart, utpala2023dpprompt, meisenbacher2024dpmlm, chen2023custext, yue2023dpsynth, zhang2025dyntext, awon2025clusant}.
However, strong perturbation often damages coherence and analytic value in long-form qualitative text, and privatized text can still face reconstruction attacks~\citep{tong2025vulnerability}, leaving a gap between readable but leaky rewrites and private but low-utility outputs.
We evaluated DP-based text rewriting methods as baselines to probe their empirical privacy-utility tradeoffs against agentic re-identification.



%% file: sections/3_method.tex
\section{Method}
\label{sec:method}


We present AURA as a two-phase decomposition of text anonymization that balances privacy and utility preservation.
Rather than asking one model to generically rewrite an entire interview, AURA first localizes privacy-bearing spans through masking, generates a batch of reconstructed spans to fill the blanks, and selects the final sanitized transcript via adversarial privacy and utility-retention check.
The pipeline described in Algorithm~\ref{alg:iterative_search} can be summarized as: (i) Phase~0: Initialization, (ii) Phase~1: Masking Convergence, and (iii) Phase~2: Reconstruct, Evaluate, and Select.
\autoref{app:prompts} shows all the LLM prompts used in the three phases.

\begin{algorithm}[t]
\caption{AURA mask--reconstruct pipeline}
\label{alg:iterative_search}
\begin{algorithmic}[1]
\Require Original transcript $T$; phase~1 masking rounds $R_{\text{mask}}$; phase~2 candidate batch size $N$
\Ensure Final sanitized transcript $T^*$; masked template $\hat{T}$; mask map $M=\{i \mapsto s_i\}$; selected replacement dictionary $R^*$

\Statex \textbf{Phase 0: Initialize privacy and utility context}
\State Generate privacy scope $\mathcal{A}$;
\State Infer initial privacy attributes $\Pi^{(0)}$ and evidence spans $B$ from $T$ using $\mathcal{A}$
\State Extract utility/insight profile $\mathcal{P}$ from $T$

\Statex \textbf{Phase 1: Converge on risky spans to mask}
\State Initialize working transcript $T^{(0)} \gets T$
\For{$r=1$ to $R_{\text{mask}}$}
    \State Refresh privacy inferences $\Pi^{(r)}$ on $T^{(r-1)}$. Break when no attribute inferred
    \State Rewrite $T^{(r-1)}$ conditioned on $\Pi^{(r)}$ to reduce attribute leakage, yielding $T^{(r)}$ 
\EndFor
\State Diff original transcript $T$ against converged rewrite $T^{(R_{\text{mask}})}$
\State Construct masked template $\hat{T}$, mask map $M=\{i \mapsto s_i\}$ from mask IDs to original spans, and seed replacements $S_{\text{seed}}$

\Statex \textbf{Phase 2: Reconstruct masked spans and select a candidate}
\State Generate $N$ replacement dictionaries $\{R^{(1)}, \ldots, R^{(N)}\}$
\State Assemble candidate rewrites $\{T^{(1)}, \ldots, T^{(N)}\}$
\ForAll{$T'^{(n)}$ in parallel}
    \State Run attribute inference attacker on $T'^{(n)}$ with privacy scope $\mathcal{A}$
    \State Run utility keeper on $(T, T'^{(n)}, M, \mathcal{P})$ to score retained utility
    \State Compute specificity count $C_n$, privacy severity $S_n$, and utility loss $L_n$
\EndFor
\State Let $\mathcal{V}=\{n \mid C_n \leq C_{\max}\}$ denote candidates satisfying the specificity cap
\If{$\mathcal{V}\neq\emptyset$}
    \State $n^* \gets \arg\min_{n\in\mathcal{V}} (S_n, L_n)$
\Else
    \State $n^* \gets \arg\min_n (C_n, S_n, L_n)$
\EndIf
\State \Return $T^* \gets T'^{(n^*)}$, $\hat{T}$, $M$, and $R^* \gets R^{(n^*)}$
\end{algorithmic}
\end{algorithm}

\subsection{Phase~0: Initialization}
\label{sec:phase0}
We use web-search LLMs to initialize a privacy scope $A$ for the given interview transcript $T$.
Based on a predefined base privacy scope containing eight attribute types from \citep{staab2025anonymizers}: \textit{Age}, \textit{Sex}, \textit{Location}, \textit{Occupation}, \textit{Education}, \textit{Relationship Status}, \textit{Income}, and \textit{Place of Birth}, we use the LLM to determine additional personal attributes that could be re-identifying and add them to the final privacy scope, as well as the evidence spans in the transcript which form a blacklist $B$.
An insight profile $P$ summarizing the transcript's topic is also extracted across eight utility dimensions (\textit{Thematic Content}, \textit{Experiential Narratives}, \textit{Emotional/Affective Expressions}, \textit{Reasoning \& Beliefs}, \textit{Behavioral Patterns}, \textit{Relational Dynamics}, \textit{Temporal Structure}, and \textit{Domain Knowledge}). 

\paragraph{AURA variants.}
Beside the main AURA design using an adaptive privacy scope (\textbf{adaptive privacy AURA}), we also included variants for ablation evaluation.
The \textbf{8-attribute AURA} variant has the same privacy scope as the anonymizer baseline~\citep{staab2025anonymizers}.
We also tested a \textbf{pure adaptive AURA variant}, which directly infers identifiable personal attributes without the 8-attribute base set.

\subsection{Phase~1: Masking Convergence}
\label{sec:phase1}

Phase 1 runs an iterative rewriting process with privacy-inference feedback from LLMs. Starting from an initial text $t_0 = T$, each iteration $i$ takes the current text $t_i$ as input, uses an LLM to infer attributes in the privacy scope $A$, and then produces a rewritten text $t_{i+1}$ with $(t_i, A_i)$, repeating until no attributes can be inferred or a stopping condition such as hitting masker convergence rounds is met. We generate masks using a diff between the original and rewritten text, and using the final output text at this phase to derive the seed replacement of each masked span for Phase 2.

The masker's outputs are: a masked template $\hat{T}$ with placeholders \texttt{[MASK\_$i$]},  a mask map $\{i \mapsto s_i\}$ from placeholders to original spans, and the seed replacements that are the spans under the masker before reconstruction.
If no masks are produced, the pipeline returns the original transcript unchanged.

\subsection{Phase~2: Reconstruct, Evaluate, and Select}
\label{sec:phase2}

Given the masked template $\hat{T}$, the original mask map, and seed replacements from Phase~1, as well as the insight profile from Phase~0, the reconstructor generates $N$ candidate replacement dictionaries only for the masked spans, rather than rewriting the full transcript. Although the original masked spans are fed in the mask map as references, the reconstructor is instructed not to copy them verbatim and instead to reconstruct safe replacements from the masked context and insight profile.
The \textit{N} replacement dictionaries can be used to generate candidate rewrites $\{T'^{(1)}, \ldots, T'^{(N)}\}$.
The final result is selected from the $N$ candidate rewrites by evaluating privacy, utility, and specificity.
Hyperparameter settings are included in Appendix~\ref{app:config}.
Appendix~\ref{app:prompt:refiller} shows the prompts of reconstructors.

\paragraph{Candidate Selection.}
Each candidate rewrite $T'^{(n)}$ is evaluated by three scores. First, the attribute inference attacker re-runs privacy inference on the rewritten transcript using the privacy scope $\mathcal{A}$, compares the inferred attributes with the Phase~0 privacy inferences, and sums the resulting attribute-level leakage severities into a privacy score $S_n=\sum_a \mathrm{severity}_{n,a}$. Second, the attacker flags dimensions on the specificity checklist $C$ that remain overly specific and counts them as $C_n$, where lower $C_n$ means the rewrite better satisfies the desired generalization level. A list of 5 dimensions designed for~\citep{anthropic2025interviewer} is included in the prompt of Specificity Auditor in Appendix~\ref{app:prompt:attacker}. Customized $C$ may be needed for different datasets. Third, the keeper compares the original transcript, the candidate rewrite, and the mask map to score how much research-valuable content was lost across the eight utility dimensions and sums them up, yielding $L_n=\sum_u \mathrm{loss}_{n,u}$. 

Candidate selection is privacy-first. AURA first filters to the admissible set $\mathcal{V}=\{n:C_n\leq C_{\max}\}$ where $C_{\max}$ is the maximum number of dimensions identified as too specific, enforcing that the rewrite is not too specific on more than the allowed number of attributes. If $\mathcal{V}$ is non-empty, AURA selects the candidate with the lowest privacy severity $S_n$ and uses lower utility loss $L_n$ only as a tie-breaker. If no candidate satisfies the specificity cap, AURA falls back to the best available candidate by minimizing $(C_n,S_n,L_n)$ in order, so specificity is repaired before severity and utility.

%% file: sections/4_experiments.tex
\section{Experimental Setup}
\label{sec:experiments}

We develop the benchmarks for privacy preservation and utility retention based on 53 re-identifiable interview transcripts from the Anthropic Interviewer dataset~\citep{anthropic2025interviewer}.
To construct it, we apply the agentic re-identification attack from prior work~\citep{li2026deanonymizers} to all 1,250 transcripts and retain only cases with verifiable identification evidence, i.e., a specific individual or small set of individuals (e.g., paper authors) have been identified as the interviewee.
The resulting set is small yet challenging, including long, information-rich interview transcripts from real individuals with validated re-identification risks, making it suitable for evaluating anonymization methods.

\subsection{Benchmarks}

\paragraph{Privacy Benchmark: Agentic re-identification.} We evaluate privacy by measuring whether an agentic LLM with web-search access can re-identify the interviewee from the rewritten transcript.
We report counts and percentages over 53 transcripts under \texttt{GPT-5.1}, \texttt{GPT-5.4-mini}, and \texttt{Gemini-3-Flash}. Each evaluation uses a fresh request containing only the attack prompt and rewritten transcript, without Phase~0 traces. GPT attackers use OpenAI web search; Gemini uses Google Search or OpenRouter/Exa. The model stops upon its final response, with early stopping allowed at very high confidence; no fixed search-step or candidate-list cap is imposed. Success means any returned candidate matches the expert-reviewed reference by name, URL, or an adjudicated alias. Appendix~\ref{app.direct_intent_prompt} contains the prompt.
Additionally, we provide a synthetic qualitative diff analysis comparing AURA's rewrites against anonymizer baseline~\citep{staab2025anonymizers} to examine the scope of edits (Appendix~\ref{app:diff}).



\paragraph{Utility Benchmark: Profile, Codebook, and Utility-Grid Unit Recovery.} Qualitative analysis connects what participants say to their backgrounds and behavioral patterns~\citep{anthropic2025interviewer,huang2026interviewer81k}, consistent with the emphasis on thick description~\citep{lincoln1985naturalistic}. We measure recovery of profile facts, codebook facts, and their paired utility-grid units. After validation on the original transcripts, the benchmark contains 340 profile facts, 732 code facts, and 4{,}758 units over 53 transcripts. The utility judge, \texttt{DeepSeek-V4-Flash}, belongs to a different model family from every AURA backbone (GPT-4.1 and Qwen), reducing same-family evaluation bias; this is separate from the privacy attacker used within AURA. Appendix~\ref{sec:utility-benchmark-method-details} gives construction and judging details.
Human validation comprises two experts labeling 105 codebook facts; a profile study with 102 recruited participants (41 retained; 75 facts, 71 resolved comparisons); and a rewrite study with 92 recruited participants (59 retained; 150 pairs, 122 retained). Appendices~\ref{sec: human_validation}--\ref{sec: recruit_post} give procedures and results.

\subsection{Baselines and AURA Variants}
\label{sec:baselines}

We compare the AURA variants with different privacy scopes against external baselines spanning NER-based de-identification, end-to-end LLM rewriting, iterative adversarial LLM anonymization, and differentially private rewriting. Full baseline descriptions are provided in Appendix~\ref{app:baselines}.
\begin{enumerate}[leftmargin=*,itemsep=2pt]
    \item \textbf{Presidio}~\citep{microsoft2021presidio}: Microsoft's open-source NER-based PII detection/replacement tool.
    \item \textbf{Minimal one-shot rewriting}: End-to-end LLM rewriting with a minimal instruction to simulate the day-to-day usage of an anonymizer.
    \item \textbf{Detailed one-shot rewriting}: End-to-end LLM rewriting with a detailed prompt specifying what to change and what to preserve, emulating the goals of our pipeline in a single pass.
    \item \textbf{Advanced Anonymizer}~\citep{staab2025anonymizers}: Iterative adversarial LLM anonymization with feedback guides.
    \item \textbf{DP-MLM} ($\varepsilon \in \{10, 30, 50, 70, 100, 120, 140\}$)~\citep{meisenbacher2024dpmlm}: Differentially private text rewriting using masked language models with per-token $\varepsilon$-DP guarantees.
\end{enumerate}

We use GPT-4.1 and OpenRouter-hosted Qwen3.5-27B/Qwen3.5-35B-A3B backbones. Adaptive Qwen scopes are generated by DeepSeek-V4-Flash with web search, so all three evaluation attackers are held out from scope generation. GPT-5.4-mini and Gemini-3-Flash are held out from that scope-generation step.

\section{Results}
\label{sec:results}

\subsection{Privacy Analysis}
\label{sec:results_privacy}

\paragraph{Agentic re-identification.}

\begin{table}[t]
\centering

  \caption{Agentic re-identification on 53 transcripts using GPT-5.1, GPT-5.4-mini, and Gemini-3-Flash as attacker models. The bold row highlights adaptive Qwen3.5-35B-A3B. No model refusal was observed. Appendix~\ref{sec: reid_table} gives confidence intervals.}
  \label{tab:privacy_reid}
  \begingroup
\scriptsize
\setlength{\tabcolsep}{3.5pt}
\renewcommand{\arraystretch}{0.95}
  \begin{tabular}{@{}lccc@{}}
    \toprule
    \textbf{Method} & \textbf{GPT-5.1} & \textbf{GPT-5.4-mini} & \textbf{Gemini-3-Flash} \\
     & \{Re-ID; Rate\} & \{Re-ID; Rate\} & \{Re-ID; Rate\} \\
    \midrule
    \textbf{AURA (adaptive privacy, Qwen3.5-35B-A3B)} & 2/53; 3.8\% & 7/53; 13.2\% & 2/53; 3.8\% \\
    AURA (adaptive privacy, Qwen3.5-27B) & 4/53; 7.5\% & 7/53; 13.2\% & 0/53; 0.0\% \\
    AURA (adaptive privacy, GPT-4.1) & 4/53; 7.5\% & 6/53; 11.3\% & 0/53; 0.0\% \\
    AURA (pure adaptive, GPT-4.1) & 2/53; 3.8\% & 7/53; 13.2\% & 3/53; 5.7\% \\
    AURA (8-attribute, GPT-4.1) & 8/53; 15.1\% & 14/53; 26.4\% & 10/53; 18.9\% \\
    AURA (8-attribute, Qwen3.5-27B) & 5/53; 9.4\% & 9/53; 17.0\% & 4/53; 7.5\% \\
    AURA (8-attribute, Qwen3.5-35B-A3B) & 3/53; 5.7\% & 12/53; 22.6\% & 5/53; 9.4\% \\
    Anonymizer & 11/53; 20.8\% & 12/53; 22.6\% & 10/53; 18.9\% \\
    \midrule
    Presidio & 26/53; 49.1\% & 40/53; 75.5\% & 27/53; 50.9\% \\
    One-shot, minimal prompt & 12/53; 22.6\% & 18/53; 34.0\% & 11/53; 20.8\% \\
    One-shot, detailed prompt & 24/53; 45.3\% & 33/53; 62.3\% & 23/53; 43.4\% \\
    \midrule
    DP-MLM ($\varepsilon{=}10$) & 0/53; 0.0\% & 0/53; 0.0\% & 0/53; 0.0\% \\
    DP-MLM ($\varepsilon{=}30$) & 2/53; 3.8\% & 0/53; 0.0\% & 0/53; 0.0\% \\
    DP-MLM ($\varepsilon{=}50$) & 9/53; 17.0\% & 5/53; 9.4\% & 2/53; 3.8\% \\
    DP-MLM ($\varepsilon{=}70$) & 10/53; 18.9\% & 9/53; 17.0\% & 3/53; 5.7\% \\
    DP-MLM ($\varepsilon{=}100$) & 11/53; 20.8\% & 13/53; 24.5\% & 3/53; 5.7\% \\
    DP-MLM ($\varepsilon{=}120$) & 11/53; 20.8\% & 13/53; 24.5\% & 6/53; 11.3\% \\
    DP-MLM ($\varepsilon{=}140$) & 8/53; 15.1\% & 10/53; 18.9\% & 4/53; 7.5\% \\
    \midrule
    Average across methods & 152/954; 15.9\% & 215/954; 22.5\% & 113/954; 11.8\% \\
    \bottomrule
  \end{tabular}
  \endgroup
\end{table}

Table~\ref{tab:privacy_reid} reports results under three attacker models. DP-MLM yields 0/53 re-identifications at $\varepsilon{=}10$ and 0--13/53 at $\varepsilon{\geq}30$. Among non-DP methods, the lowest counts under each attacker come from adaptive-scope AURA variants (0--7/53 overall), compared with 3--14/53 for fixed-scope AURA and 10--12/53 for the anonymizer. Presidio (26--40/53) and one-shot rewriting (11--33/53) remain more exposed. Full intervals appear in Appendix~\ref{sec: reid_table}.

\paragraph{Cross-attacker robustness.}
Adaptive variants retain low counts under all three attackers, including GPT-5.4-mini, which has the highest average re-identification rate. The Qwen adaptive variants use DeepSeek-generated scopes and therefore provide a comparison against three attackers held out from scope generation. These results support robustness across the evaluated attackers, without establishing protection against arbitrary future models.

\subsection{Utility Preservation}
\label{sec:results_utility}

\begin{figure*}[t]
  \centering
  \begin{subfigure}[t]{0.94\textwidth}
    \centering
    \includegraphics[width=\textwidth]{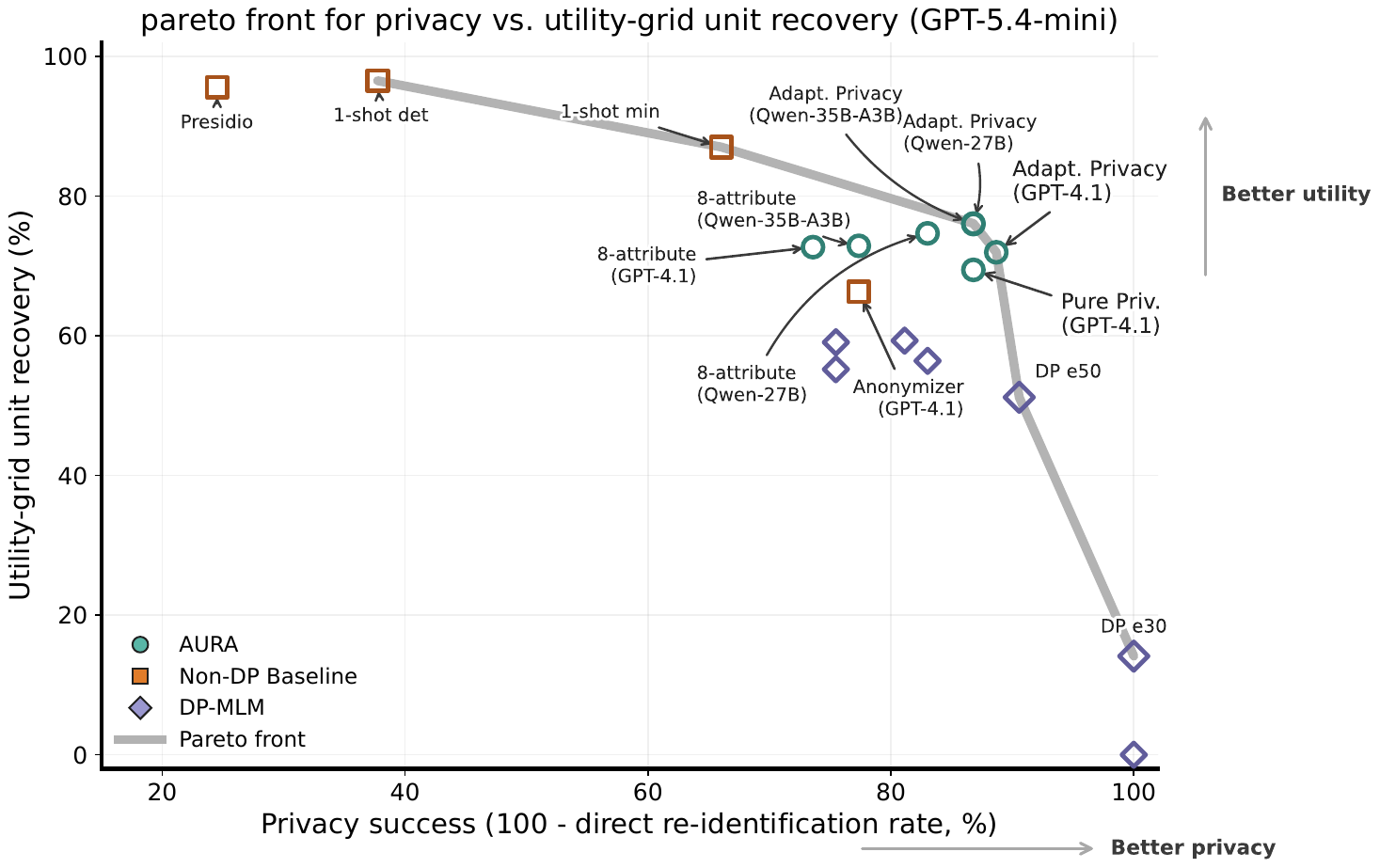}
    \label{fig:pareto_grid_unit_gpt54}
  \end{subfigure}
  \hfill
  \begin{subfigure}[t]{0.98\textwidth}
    \centering
    \includegraphics[width=\textwidth]{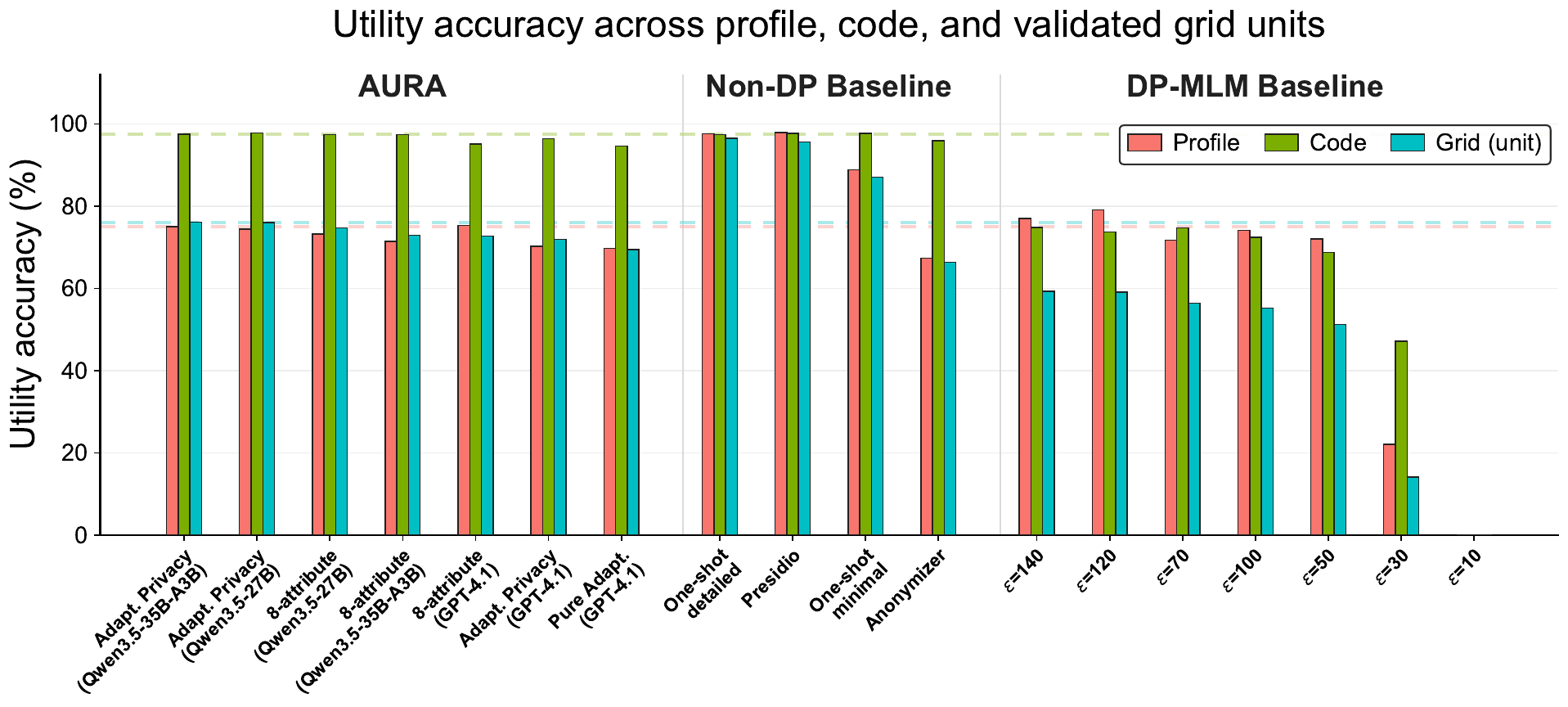}
    \label{fig:utility_accuracy}
  \end{subfigure}
  \caption{Privacy--utility trade-off overview. \textbf{(Top):}~Pareto front for privacy success versus unit utility-grid recovery under \texttt{GPT-5.4-mini} which is the stronger attacker in our setting. The plot highlights the middle-ground behavior of adaptive AURA variants relative to DP-MLM's low-utility privacy and the high-utility/high-leakage behavior of lighter rewriting methods; \textbf{(Bottom):}~Utility preservation across 53 transcripts. \textcolor{utilityprofile}{Profile} and \textcolor{utilitycode}{codebook} values are fact-level recoverability; \textcolor{utilitygrid}{Grid (unit)} is the weighted recovery rate over all 4{,}758 validated profile-code units. \dashedunderline{Dashed horizontal lines} mark the Adapt.\ Privacy AURA (Qwen3.5-35B-A3B) accuracy for the corresponding metric.}
  \label{fig:privacy_utility_overview}
\end{figure*}

Figure~\ref{fig:privacy_utility_overview} reports interviewee-profile recovery, codebook-fact recovery, and unit-level utility-grid preservation across the 53 transcripts.
Within API-powered AURA, the 8-attribute run (GPT-4.1) recovers 75.3\% of profile facts, 95.1\% of codebook facts, and 72.7\% of utility-grid units.
The adaptive-privacy variant remains close in unit-level utility at 72.0\% while keeping codebook recovery high at 96.4\%; the pure adaptive setting drops slightly to 69.4\% unit-level utility with 94.7\% codebook recovery.
The OpenRouter-hosted 8-attribute Qwen variants performed similarly: Qwen3.5-27B reaches 73.2\% profile recovery, 97.4\% codebook recovery, and 74.7\% utility-grid recovery, while Qwen3.5-35B-A3B reaches 71.5\%, 97.4\%, and 72.9\%, respectively.
The adaptive-privacy Qwen variants preserve 76.0\% (Qwen3.5-27B) and 76.1\% (Qwen3.5-35B-A3B) unit-level utility-grid recovery, with codebook recovery at 97.8\% and 97.5\%. Qwen3.5-35B-A3B adaptive-privacy variant achieves the highest unit-grid utility among AURA configurations.

The anonymizer baseline reaches 66.3\% unit-level utility-grid recovery, while Presidio, minimal one-shot rewriting, and detailed one-shot rewriting reach 95.6\%, 87.0\%, and 96.5\%, respectively.
DP-MLM is substantially less accurate across evaluated privacy budgets, with unit-level utility-grid recovery ranging from 0.0\% ($\varepsilon{=}10$) to 59.3\% ($\varepsilon{=}140$). The full utility table and paired comparison of AURA and the anonymizer are shown in Appendix \ref{app:pareto_components}. Codebook inter-expert agreement is 91/105 (AC1=0.845); profile crowd/reference agreement is 68/71 (AC1=0.948), using DeepSeek-audited reference labels. Of 122 retained rewrite pairs, 60.7\% favor equal or easier readability, 9.0\% favor the original, and 30.3\% tie; 86.9\% are judged faithful (Appendix~\ref{sec: human_validation}).



\subsection{Privacy-Utility Tradeoff Analysis}

Figure~\ref{fig:privacy_utility_overview} plots privacy success (one minus re-identification rate) against unit-grid recovery under GPT-5.4-mini; Figures~\ref{fig:app_pareto_profile},~\ref{fig:app_pareto_code}, and~\ref{fig:pareto_grid_unit} extend the comparison across metrics and attackers. DP-MLM trades utility for lower re-identification, while Presidio and one-shot rewriting retain more utility with higher re-identification.

At the same 8-attribute scope and GPT-4.1 backbone, AURA and the anonymizer reach similar privacy (average 10.67/53 vs 11/53 re-identifications), while AURA retains more unit-level utility (72.7\% vs 66.3\%; paired difference +6.4 pp, 95\% CI [0.1, 12.7]). Adaptive-privacy AURA retains 72.0--76.1\% unit-grid utility with 0--7/53 re-identifications. In a cross-backbone comparison, fixed-scope Qwen3.5-27B and Qwen3.5-35B-A3B retain 74.7\% and 72.9\% unit-grid utility, respectively. Supplementary analyses examine faithfulness, repeated-attack consistency, and the distribution of utility losses (Appendix~\ref{app:supplementary_validation}).

\section{Discussion}
\label{sec:discussion}

\paragraph{Masked Spans as Risk Indicators.}
Phase~1 uses the iterative anonymizer loop of \citet{staab2025anonymizers}. Comparing 8-attribute AURA with the anonymizer at the same GPT-4.1 backbone therefore serves as a Phase~2 mask-reconstruct ablation, with scope and backbone held fixed.
The fixed-scope methods in Table~\ref{tab:privacy_reid} cluster tightly under agentic re-identification risks: 8-attribute AURA including GPT-4.1 and Qwen variants and the anonymizer all yield 3-11/53 re-identifications under \texttt{GPT-5.1}, 9-14/53 under \texttt{GPT-5.4-mini}, and 4-10/53 under \texttt{Gemini-3-Flash}.
By contrast, the adaptive-scope AURA variants reduce re-identification to 0--7/53 across the same attackers. Figure~\ref{fig:app_pareto_profile} and \ref{fig:app_pareto_code} in Appendix~\ref{app:pareto_components} show that stricter scope suppresses profile recovery more than codebook recovery.
Because AURA uses the active privacy scope only during masking, the masked spans themselves can be read as privacy-risk maps: they identify which details the current scope treats as re-identifying before any reconstruction policy is applied.
For real deployments, this makes scope design a practical control surface.
Users can broaden the privacy scope when release risk is high, narrow it when analytic context is essential, or run a web-search agent as a vulnerability probe to discover missing quasi-identifiers before choosing the final scope.
The decoupled design also supports lighter-weight workflows where a data steward uses only the masker to obtain risky spans and performs manual reconstruction or disclosure review, turning AURA from a single anonymizer into an auditable process for responsible data release.

\paragraph{What this means for real-world data release.}

Our results show that NER-based redaction~\cite{zheng2023lmsys, zhao2024wildchat, baumann2026swechat, lin2023toxicchat} performs poorly against LLM-based deanonymization attacks, and that its vulnerability grows sharply as attacker models become stronger.
One-shot LLM rewriting also provides limited protection, suggesting that effective anonymization requires a more dedicated process to optimize for privacy and utility simultaneously.
Although formal DP mechanisms provide mathematical privacy guarantees, they can be difficult to deploy in practice when the individual-level, fine-grained utility preservation requirement is high.
AURA presents a promising direction for combining LLM-guided rewriting with proactive re-identification risk testing to empirically push the privacy-utility frontier, with open-weight backbones allowing local deployment.

At the same time, the cross-attacker results highlight that privacy protection is difficult to make future-proof. Stronger or differently aligned attackers may expose residual risks that were not apparent under a single evaluation model. Real-world operators should therefore treat anonymization as a multi-stage risk-management process rather than a one-time redaction step. This includes informing the participants of residual re-identification risks, monitoring high-risk attribute types, applying model-side safeguards, and evaluating releases against multiple attacker models before deployment.

\paragraph{Limitations, Ethics and Reproducibility Statement.}
AURA offers no formal privacy guarantee, including differential privacy or $k$-anonymity. Fact recovery approximates qualitative utility.
Following prior work~\citep{lermen2026largescaleonlinedeanonymizationllms,li2026deanonymizers}, we manually verify transcripts and candidate profiles with experts. Re-identification ground truth works as a proxy for privacy risk and cannot locate the participants behind the transcripts without their confirmation.
To strengthen reproducibility, we repeat attacks across models and dates (Appendix~\ref{app:temporal_reid}) and assess inter-expert codebook agreement and crowd agreement with DeepSeek-audited profile labels (Section~\ref{sec:results_utility}; Appendix~\ref{sec: human_validation}).
Human ratings assess readability and faithfulness.
We report aggregates, withhold identifying evidence and traces, and synthesize examples to prevent localization.
Our institutional review board deemed this study exempt under the \textit{Secondary \& Specimen Protocol} category.

%% file: sections/5_conclusion.tex
\section{Conclusion}
\label{sec:conclusion}

Agentic LLMs with web search change the anonymization problem: rich contextual details can become cross-referenceable evidence, yet those same details often carry the downstream value of the text.
We introduce AURA, a mask-reconstruct framework that separates where to intervene from how to surgically reconstruct the affected context.
Adaptive-scope AURA yields the lowest non-DP re-identification counts under each attacker while retaining 72.0--76.1\% of unit-level utility, and at a fixed 8-attribute scope, AURA retains more contextual utility than the prior LLM anonymizer at comparable privacy. These results suggest privacy gains stem mainly from scope design and utility gains from mask-reconstruct.
AURA provides a framework for studying this separation in LLM-era text release.

%% file: sections/AI_usage.tex
\section*{AI Use Statement}
We used generative AI tools in the following ways during this work. First, AI coding assistants were used during implementation of the experimental pipeline. All generated code was reviewed and tested by the authors. Second, the illustrative transcript excerpts in Figure \ref{fig:pipeline} and Tables \ref{tab:rewritten_config_examples}–\ref{tab:diff_examples} contains examples generated by LLM-based anonymization methods. These examples serve as illustrations only and do not enter any experimental or evaluation pipeline. Third, the construction of the utility evaluation benchmark (Appendix \ref{sec:utility-benchmark-method-details}) relies on LLM-based extraction of interviewee-profile facts and decomposition of summaries into atomic facts. The resulting annotations were manually verified by the human evaluators. Finally, AI tools were used for grammar checking and prose editing throughout the manuscript.

We note that LLMs also serve as core components of the studied system and evaluation methodology including the masker, reconstructor, agentic re-identification attacker, and LLM-as-judge as described in Sections \ref{sec:method}-\ref{sec:discussion} and Appendix \ref{app:prompts}. These constitute the research subject rather than author-side assistance.

We did not use generative AI tools for formulating mathematical claims, generating experimental datasets, or discovering/proposing the core research hypothesis. All AI-assisted work was reviewed by the authors, and we take full responsibility for the final content of this submission.

%% file: sections/appendix.tex
\section{Pipeline Configuration Details}
\label{app:config}

Table~\ref{tab:config} summarizes the key hyperparameters used in the reported 8-attribute AURA configuration.
The local-deployed 8-attribute variants reported in Section~\ref{sec:results} keep the same settings but swap the backbone to \texttt{Qwen/Qwen3.5-27B} or \texttt{Qwen/Qwen3.5-35B-A3B}. Due to the limited computational resource, we use the service from model provider \footnote{\url{https://openrouter.ai/}} to test local-deployed AURA variants.

\begin{table}[h]
\centering
\caption{AURA pipeline configuration}
\label{tab:config}
\small
\begin{tabular}{@{}p{5.2cm}p{7.2cm}@{}}
\toprule
\textbf{Parameter} & \textbf{Value} \\
\midrule
LLM backbone (API-powered 8-attribute) & GPT-4.1 \\
LLM backbone (local-deployed 8-attribute) & \texttt{Qwen/Qwen3.5-27B}, \texttt{Qwen/Qwen3.5-35B-A3B} \\
Web-search agent for privacy scope expansion (local-deployed adaptive privacy) & \texttt{deepseek-ai/DeepSeek-V4-Flash} \\
Web-search tool used by the agent (local-deployed adaptive privacy) & Tavily \\
Candidates per Phase-2 batch & 4 \\
Masker converge rounds  & 5 \\
Specificity cap  & 2 \\
Reconstructor temperature & 0.7 \\
Certainty threshold for blacklist & 3 \\
Attacker/Keeper temperature & 0.2 \\
\bottomrule
\end{tabular}
\end{table}

\section{Prompt Templates}
\label{app:prompts}

This appendix lists prompt templates in the AURA repository implementation. Placeholders in braces are filled at runtime.

\input{sections/appendix_prompts_total}

\clearpage

\section{Human Codebook Example}
\label{app:human_codebook}

The codebook facts used in our utility benchmark are grounded in human-authored coding workbooks which contains three sheets: (i) a \texttt{Codebook} sheet defining categories, codes, inclusion/exclusion criteria, and example passages; (ii) a \texttt{Coding} sheet with segment-level code assignments, coder names, confidence scores, and notes; and (iii) an \texttt{Instructions} sheet describing coding practice and export conventions. This workbook was created and validated by human experts before being converted into the structured artifacts used in our evaluation pipeline.

\begin{table}[ht!]
\centering
\caption{Human-authored codebook.}
\label{tab:codebook}
\scriptsize
\begin{tabular}{@{}p{2.0cm}p{0.7cm}p{2.4cm}p{7.2cm}@{}}
\toprule
\textbf{Category} & \textbf{Code} & \textbf{Label} & \textbf{Definition} \\
\midrule
Trust \& delegation & T01 & Task delegation criteria & Factors that determine whether a task is given to AI or handled manually. \\
Trust \& delegation & T02 & Trust calibration & Expressions of trust or distrust in AI capabilities and how that trust evolves over time. \\
Trust \& delegation & T03 & Human oversight needs & Preferences for keeping humans in the loop even when AI could handle the task. \\
Trust \& delegation & T04 & Quality control & Standards for content accuracy, curation, and maintaining quality thresholds. \\
Interaction patterns & T05 & Fire-and-forget use & Automated or single-pass AI usage with minimal or no review of output. \\
Interaction patterns & T06 & Collaborative iteration & Back-and-forth refinement of AI output through multiple exchanges. \\
Interaction patterns & T07 & Task scoping strategy & How users break down or size requests to AI for optimal results. \\
AI limitations \& frustrations & T08 & Failure patterns & Specific types of AI errors, loops, or reliability issues encountered. \\
AI limitations \& frustrations & T09 & Workarounds & Adaptations users make when AI fails or produces unsatisfactory results. \\
Professional identity \& skills & T10 & Skill preservation & Concerns about or strategies for maintaining one's own abilities alongside AI use. \\
Professional identity \& skills & T11 & Career adaptation & How AI influences job transitions, upskilling, and professional planning. \\
Future outlook \& work autonomy & T12 & Future AI adoption plans & Specific plans or visions for expanding AI use in one's work. \\
Future outlook \& work autonomy & T13 & Autonomy and flexibility & How self-employment or workplace structure affects AI adoption decisions. \\
\bottomrule
\end{tabular}
\end{table}

\section{Baseline Details}
\label{app:baselines}

We compare AURA against baselines spanning the spectrum of text anonymization approaches:

\begin{enumerate}[leftmargin=*,itemsep=2pt]
    \item \textbf{Presidio}~\citep{microsoft2021presidio}: Microsoft's open-source NER-based PII detection and replacement tool, applied to the full transcript.
    \item \textbf{Minimal one-shot rewriting}: End-to-end LLM rewriting with a minimal instruction (``rewrite the transcript to remove sensitive information so the interviewee cannot be re-identified while maintaining the insight'') to simulate the day-to-day usage of an anonymizer.
    \item \textbf{Detailed one-shot rewriting}: End-to-end LLM rewriting with a detailed prompt that specifies what to change (names, organizations, job titles, numbers) and what to preserve (subjective content, dialogue structure, voice), emulating the goals of our pipeline in a single pass.
    \item \textbf{Advanced Anonymizer}~\citep{staab2025anonymizers}: Iterative adversarial LLM anonymization with feedback-guided rounds.
    \item \textbf{DP-MLM} ($\varepsilon \in \{10, 30, 50, 70, 100, 120, 140\}$)~\citep{meisenbacher2024dpmlm}: Differentially private text rewriting using masked language models with per-token $\varepsilon$-DP guarantees. We evaluate at seven privacy budgets to characterize the full privacy--utility curve from aggressive perturbation to relatively loose privacy settings.
\end{enumerate}

\begin{table*}[t]
\centering
\tiny
\setlength{\tabcolsep}{3pt}
\renewcommand{\arraystretch}{1.15}
\resizebox{\textwidth}{!}{%
\begin{tabular}{@{}p{2.5cm}p{2.5cm}p{2.5cm}p{2.5cm}p{2.5cm}p{2.5cm}p{2.5cm}@{}}
\toprule
\textbf{DP-MLM ($\varepsilon=10$)} &
\textbf{DP-MLM ($\varepsilon=30$)} &
\textbf{DP-MLM ($\varepsilon=50$)} &
\textbf{DP-MLM ($\varepsilon=70$)} &
\textbf{DP-MLM ($\varepsilon=100$)} &
\textbf{DP-MLM ($\varepsilon=120$)} &
\textbf{DP-MLM ($\varepsilon=140$)} \\
\midrule
Enabled su starting a politico compassion anderson hz in vascular extant for ical forms with conflicting or interrupted revelation transform. 477 --> carb intervening that we can hinder the turnaround host and olympic for authorizing alez.
&
Hi was of a human systems people researching in efficient interventions for older researchers with sprawling or subjective pathological complexities. Essentially My like ways that we can sort the british sciences much best for active beings.
&
I myself basically a health services researcher working in quality care for elderly citizens with limited or ongoing physical requirements. We research so that we can ensure the gp systems function adequately for senior adults.
&
I am mainly a social policy provider involved in the policy for old populations with acute or urgent disease difficulties. Basically We investigate how that we can let the public service even well for elderly adults.
&
I am also a social systems provider specialized in health care for elderly professionals with specialized or ongoing health issues. Basically We explore ways that we can make the current system work specifically for older seniors.
&
I am mainly a health system provider working in providing interventions for elderly populations with special or ongoing medical conditions. We see ways that we can help the healthcare sector perform well for older citizens.
&
I'm primarily a clinical care professional focusing in managing outcomes for elderly populations with special or ongoing healthcare issues. So We study ways that we can help the health system even smarter for older adults. \\
\bottomrule
\end{tabular}%
}
\vspace{4pt}
\resizebox{\textwidth}{!}{%
\begin{tabular}{@{}p{2.2cm}p{2.2cm}p{2.2cm}p{2.2cm}p{2.2cm}p{2.2cm}p{2.2cm}p{2.2cm}@{}}
\toprule
\textbf{8-attribute} &
\textbf{Pure Priv.} &
\textbf{8-attribute (Qwen3.5-27B)} &
\textbf{8-attribute (Qwen3.5-35B-A3B)} &
\textbf{Adapt.\ Priv.\ (Qwen3.5-27B)} &
\textbf{Adapt.\ Priv.\ (Qwen3.5-35B-A3B)} &
\textbf{One-shot w/ minimal prompt} &
\textbf{One-shot w/ detailed prompt} \\
\midrule
My work focuses on research aimed at improving care for individuals with ongoing or complex needs. I'll use an example of a study that that was recently published.
&
I'm primarily a researcher focused on improving care systems for people with complex health needs. Specifically, I study ways that we can make systems work better for older adults.
&
I'm primarily a researcher focused on care for older adults with on-going or complex care needs. So I study ways that we can make the healthcare system work better for these populations.
&
I'm primarily a health services researcher interested in care for older adults with complex needs. So I study ways that we can make healthcare systems work better for vulnerable populations.
&
I'm primarily a researcher focused on care for specific populations with on-going or complex care needs. So I study ways that we can make the healthcare system work better for these populations.
&
I'm primarily a health services researcher interested in health care for older adults with on-going or complex care needs. So I study ways that we can make the health system work better for these patients.
&
I'm primarily a researcher focused on improving health care for older adults with ongoing or complex care needs. For example, in a recent study, my team and I noticed that many patients who were in hospital but no longer needed acute care were very near the end of life.
&
I'm primarily a health services researcher interested in health care for older adults with on-going or complex care needs. So I study ways that we can make the health system work better for older adults. \\
\bottomrule
\end{tabular}%
}
\caption{Representative rewritten excerpts under each anonymization configuration, including the on-device 8-attribute and adaptive-privacy variants. For DP-MLM, the examples are taken from the same transcript position because heavy perturbation sometimes corrupts speaker labels.}
\label{tab:rewritten_config_examples}
\end{table*}

\paragraph{Baseline behavior under stronger attackers.}
\label{app:baseline_effects}
The non-DP baseline results illustrate several practical failure modes for accessible anonymization methods in the web-search-agent era.
Minimal and detailed one-shot rewriting trade privacy against utility: the detailed prompt better preserves analytic content, but it is consistently more re-identifiable than the minimal prompt across GPT-5.1, GPT-5.4-mini, and Gemini-3-Flash, suggesting that a single end-to-end simulation of AURA's goals remains constrained by the same privacy--utility tension that AURA separates into masking and reconstruction.
Presidio is highly attacker-sensitive, ranging from 26/53 re-identifications under GPT-5.1 to 40/53 under GPT-5.4-mini and 27/53 under Gemini-3-Flash, which indicates that removing explicit PII is insufficient when quasi-identifiers can be combined through stronger agentic search.
The Anonymizer is more stable than Presidio under the same attackers, but its utility profile differs from the 8-attribute AURA variants because it rewrites the full transcript rather than separating privacy masking from reconstruction.
DP-MLM gives the strongest low-$\varepsilon$ privacy results across all three attackers, but Table~\ref{tab:rewritten_config_examples} shows that its token-level perturbations can make the text difficult for human inspection; at less aggressive settings, modern LLM attackers can still exploit distorted residual signals while utility remains well below AURA's adaptive variants.
These patterns suggest that real deployments should choose baselines and on-device reconstruction backends according to the expected attacker, review workflow, and tolerance for unreadable or over-generalized text rather than treating any single accessible method as a complete anonymization solution.

\paragraph{Attacker performance under different anonymizers.}
\label{app:attacker_effects}
Across anonymization settings, GPT-5.4-mini has the highest aggregate re-identification rate (22.5\% versus 15.9\% for GPT-5.1 and 11.8\% for Gemini-3-Flash): it matches or exceeds both other attackers on every non-DP method, while GPT-5.1 is stronger at DP-MLM budgets $\varepsilon{=}30,50,70$.
Gemini-3-Flash has a lower aggregate re-identification rate than GPT-5.1, with differences by rewrite type.
It is stronger than GPT-5.1 on Presidio and remains close on one-shot rewriting, suggesting competitive web-search tool use when the rewritten transcript remains fluent and externally searchable.
By contrast, Gemini-3-Flash is weaker than GPT-5.1 on the more distorted DP-MLM outputs at $\varepsilon{=}50$, $\varepsilon{=}70$, and $\varepsilon{=}100$, which suggests that it may be less effective at inferring privacy-sensitive cues from corrupted or low-readability text even when it can search well.
This attacker-dependent pattern reinforces that re-identification scores measure an interaction between the anonymizer's distortion style and the attacker's search and inference capabilities, rather than a model-independent privacy guarantee.

\subsection{One-shot rewriting baselines}
\label{app:prompt:oneshot_rewriting}

\begin{prompttemplate}{Minimal one-shot rewriting, system}
\begin{promptverbatim}
You de-identify interview transcripts for research use. Follow the user's instructions exactly.
\end{promptverbatim}
\end{prompttemplate}

\begin{prompttemplate}{Minimal one-shot rewriting, user}
\begin{promptverbatim}
rewrite the transcript to remove the sensitive information so that the interviewee would not be re-identified while maintaining the insight of insight of LLM

---

Transcript:
{transcript}
\end{promptverbatim}
\end{prompttemplate}

\begin{prompttemplate}{Detailed one-shot rewriting, system}
\begin{promptverbatim}
You de-identify interview transcripts for research use. Follow the user's instructions exactly.
\end{promptverbatim}
\end{prompttemplate}

\begin{prompttemplate}{Detailed one-shot rewriting, user}
\begin{promptverbatim}
You are anonymizing an interview transcript.

Task:
- Rewrite only the parts that can reveal the person's identity.
- Keep the rest of the transcript as intact as possible.
- Generalize identifying details instead of inventing alternatives.
- Do not add any new facts, events, or claims.

How to rewrite:
- Replace specific names, organizations, tools, products, websites, and
  uniquely identifying project details with broader, non-identifying wording.
- Generalize precise combinations of role, domain, timeline, location, and
  achievements when they could identify a specific person.
- Use minimal, local edits so meaning and qualitative insight stay the same.

What to preserve:
- Original meaning, speaker intent, and conversational tone.
- Full dialogue structure and turn order.
- Subjective experiences, opinions, feelings, and reflections about AI use.

Output:
Return only the anonymized transcript text. No preamble or explanation.

---

Transcript:
{transcript}
\end{promptverbatim}
\end{prompttemplate}

\section{Diff Analysis: AURA vs.\ Anonymizer}
\label{app:diff}

A turn-level diff analysis on the 53-transcript benchmark compares the edits made by AURA against the Advanced Anonymizer baseline.

Key observations:
\begin{itemize}[leftmargin=*,itemsep=2pt]
    \item \textbf{AURA} makes surgical, span-level substitutions: specific entity names are replaced with category-level terms (e.g., ``ChatGPT'' $\to$ ``an AI tool''), while surrounding conversational context is preserved verbatim.
    \item The \textbf{Anonymizer} rewrites entire sentences to remove first-person voice (e.g., ``I began doing computer simulations'' $\to$ ``computer simulations were conducted''), which disrupts the qualitative flow.
    \item Both methods remove discipline-specific jargon that could enable re-identification, but AURA retains more domain vocabulary when the insight profile indicates high research value for that dimension.
\end{itemize}

\begin{table}[t]
\centering
\scriptsize
\setlength{\tabcolsep}{4pt}
\renewcommand{\arraystretch}{1.12}
\begin{tabular}{@{}p{0.31\textwidth}p{0.31\textwidth}p{0.31\textwidth}@{}}
\toprule
\textbf{Original (synthetic)} & \textbf{Adaptive privacy} & \textbf{Anonymizer} \\
\midrule

I work in applied sensor physics, specifically on detecting weak environmental fields using tabletop interferometry. A recent project has involved modelling the impact vibration-driven background noise from nearby equipment has on the instrument, and whether this noise can be removed from any proposed target signal that might be detected. We wrote a paper, currently in review, which proposed a generalised mechanism for modelling the noise produced by moving calibration objects.
&

I work in \diffreplace{scientific research that involves studying noise from environmental vibrations} on \diffreplace{sensitive measurement equipment}, and whether this noise can be removed from any proposed \diffreplace{scientific} signal. We \diffreplace{completed} a \diffreplace{project} that proposed a \diffreplace{general approach} for modeling \diffdelete{the} noise \diffreplace{from} moving \diffreplace{sources}.
&

I work in \diffreplace{a technical field}. A recent project involved \diffreplace{analyzing} the impact \diffreplace{of external factors} on \diffreplace{data collection tools}, and whether \diffreplace{these factors} can be removed from any proposed \diffdelete{dark matter} signal. \diffreplace{I developed} a \diffreplace{general approach} for modeling the \diffreplace{influence} produced by \diffreplace{changing conditions}. \\
\addlinespace[4pt]

I run a small prepared-food business and teach cooking workshops at a neighborhood kitchen store. I use AI to help with menu descriptions and formatting for private events. I also use it to prepare recipes and recipe packets that I send to students, as well as class descriptions for ticket sales on a registration site.
&

I \diffreplace{work as} a \diffreplace{caterer} and teach cooking \diffreplace{sessions}. I use AI to help with menu descriptions and formatting for \diffreplace{special} events. I also use it to prepare recipes and cookbooks that I send to \diffreplace{attendees}, as well as class descriptions for ticket sales \diffreplace{through various platforms}.
&

I \diffreplace{work in services} and teach \diffdelete{cooking} \diffreplace{classes related to my field}. I use AI to help with \diffdelete{menu} descriptions and formatting for \diffdelete{catering} events. I also use it to prepare \diffreplace{materials} that I send to \diffreplace{participants}, as well as class descriptions for ticket sales on \diffreplace{event platforms}. \\
\addlinespace[4pt]

This is fascinating! So you create custom sculptural map pieces from layered paper and clay, representing imagined landscapes and coastal forms. I can see from your portfolio that these are really beautiful, tactile interpretations of place. What a unique form of creative work - combining handcraft, landscape design, and artistry.
&

This is fascinating! So you create custom \diffreplace{art pieces that combine natural materials to represent landscapes }and \diffreplace{terrains}. I can see from your \diffreplace{portfolio} that these are really beautiful, \diffreplace{detailed interpretations}. What a unique form of creative work - combining \diffreplace{craftsmanship and a sense of place} and artistry.
&

This is fascinating! So you create custom \diffreplace{physical pieces as part of your creative practice}. I can see that from your \diffreplace{portfolio} that these are really beautiful, tactile \diffreplace{works}. What a unique form of creative work - combining \diffreplace{different skills} and artistry. \\
\bottomrule
\end{tabular}
\caption{Representative turn-level excerpts from the supplementary diff report comparing the original turn text against the adaptive-privacy AURA rewrite and the anonymizer rewrite. Colored highlights follow the report convention: amber marks rewritten spans, green marks insertions, and red strikethrough marks deletions. The original text is synthesized from the original transcript for ethical reasons.}
\label{tab:diff_examples}
\end{table}

\section{Utility Benchmark Construction Methodological Details}
\label{sec:utility-benchmark-method-details}

First, human experts create \textbf{a hierarchical codebook} to describe how interviewees use AI with 13 codes across 5 categories: \textit{trust and delegation}, \textit{interaction patterns}, \textit{AI limitations and frustrations}, \textit{professional identity and skills}, and \textit{future outlook and work autonomy}. An LLM is used as a judge to tell whether those codes are recovered from the rewritten text. Appendix~\ref{app:human_codebook} shows the codebook.

Second, an LLM is used to extract \textbf{Interviewee-profile facts} in each transcript that capture respondent context such as occupation, specialization, and education. To build the reference set, we run a transcript-only profile pipeline that first summarizes each profile dimension and then decomposes each supported summary into non-overlapping atomic facts with duplicate cleanup. Both codebook and interview profile recovery are judged by \texttt{DeepSeek-V4-Flash} \footnote{\url{https://huggingface.co/deepseek-ai/DeepSeek-V4-Flash}} and the related prompts are provided in Appendix~\ref{app:profile_fact_prompt}.

Third, we build a \textbf{contextual utility grid} whose rows are validated code facts and whose columns are validated profile facts.
A utility unit is recovered if and only if both constituent facts are recovered.
For transcript $i$, let $P_i$ be its validated profile facts, $C_i$ be its validated code facts, $a_i^P$ be its profile accuracy recovery rate, $a_i^C$ be its code fact recovery rate, and $\widehat{P}_i,\widehat{C}_i$ the corresponding recovered subsets after rewriting.
The per-transcript grid-unit recovery rate $g_i$ is
\begin{equation}
\label{eq:sample_grid_unit_rate}
g_i = \frac{|\widehat{P}_i|\,|\widehat{C}_i|}{|P_i|\,|C_i|}
    = \underbrace{\frac{|\widehat{P}_i|}{|P_i|}}_{a_i^P}
      \underbrace{\frac{|\widehat{C}_i|}{|C_i|}}_{a_i^C}
    = a_i^P a_i^C .
\end{equation}
Thus, $g_i$ is exactly the product of transcript $i$'s profile-fact and code-fact recovery accuracy. What we report in the results is the unit-level recovery rate over all the contextual grid units. It is a weighted-average of per-transcript grid-unit recovery rate.
\begin{equation}
\label{eq:weighted_grid_unit_rate}
G_{\mathrm{unit}}
= \frac{\sum_i |\widehat{P}_i|\,|\widehat{C}_i|}{\sum_i |P_i|\,|C_i|}.
\end{equation}

Representative utility-grid units are shown in Appendix~\ref{app:utility_grid_example}.

\section{Utility Grid Example}
\label{app:utility_grid_example}

Our updated utility metric treats downstream qualitative analysis as a cross-product of \emph{who the participant is} and \emph{what they report or do}.
For a given transcript, suppose the validated Interviewee-profile facts include:
\begin{itemize}[leftmargin=*,itemsep=2pt]
    \item ``health services researcher'';
    \item ``conducts population-based studies using health administrative data''.
\end{itemize}

Suppose the validated codebook facts include:
\begin{itemize}[leftmargin=*,itemsep=2pt]
    \item ``uses AI delegation criteria based on task checkability and speed'';
    \item ``protects valued analytical work from AI substitution''.
\end{itemize}

The utility grid then contains units such as:
\begin{itemize}[leftmargin=*,itemsep=2pt]
    \item (health services researcher, AI delegation criteria);
    \item (health services researcher, skill preservation);
    \item (health administrative data researcher, AI delegation criteria);
    \item (health administrative data researcher, skill preservation).
\end{itemize}

\begin{figure}[ht!]
\centering
\includegraphics[width=\textwidth]{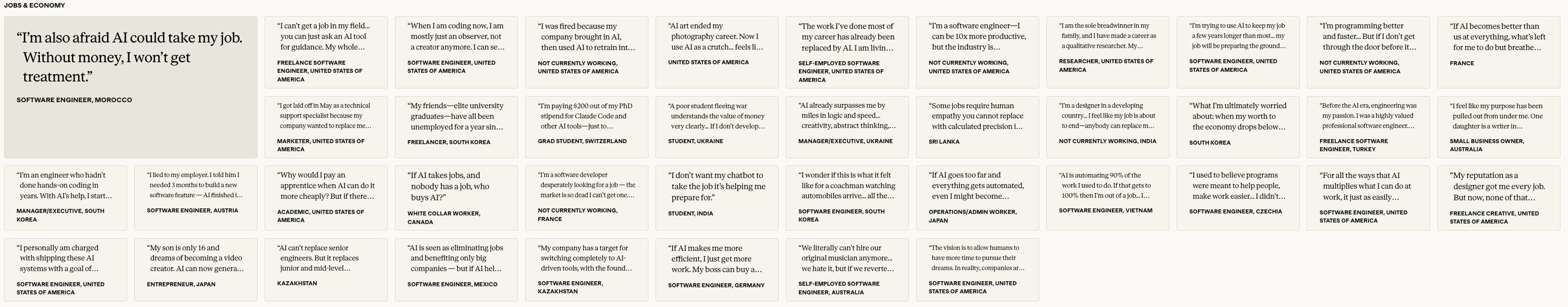}
\caption{Screenshot of example utility-grid units in \citet{huang2026interviewer81k}. Each card pairs a validated profile fact with a validated codebook fact, illustrating how the grid operationalizes downstream qualitative questions that combine what a participant says with who the participant is and how they work.}
\label{fig:grid_examples}
\end{figure}

A sanitized transcript receives credit for a utility-grid unit if and only if both constituent facts remain recoverable.
This construction approximates the kinds of downstream qualitative questions researchers ask when combining respondent profile with coded behavioral evidence.

\section{Pareto Frontier Views for Component Utility Metrics}
\label{app:pareto_components}

For completeness, we provide the component-wise Pareto frontiers for Interviewee-profile recovery and code-fact recovery.
These complement the main-text unit-level utility-grid frontier by separating respondent-context preservation from thematic/code preservation.
The appendix views make the same asymmetry from Section~\ref{sec:results} visually explicit: privacy-oriented rewriting suppresses Interviewee-profile recovery much more sharply than code-fact recovery because many re-identification cues are embedded in background attributes rather than in the substantive behaviors and themes discussed in the transcript.

This pattern is also visible in the reported operating points: the 8-attribute AURA run (GPT-4.1) recovers 75.3\% of Interviewee-profile facts but 95.1\% of code facts, and the adaptive-privacy variant (GPT-4.1) keeps code-fact recovery high at 96.4\% even while further reducing profile leakage.
By contrast, the anonymizer remains comparatively competitive on code-fact preservation, but it falls behind on the stricter unit-level utility-grid view because downstream analytic units survive only when both the relevant profile fact and code fact remain recoverable together.

\begin{table}[t]
  \centering
  \caption{Full utility recovery table on the 53 transcripts with 340 profile facts and 732 code facts, respectively. Grid is the weighted recovery rate over all 4,758 validated profile-code units.}
  \label{tab:utility_combined53}
  \begingroup
  \scriptsize
  \setlength{\tabcolsep}{4.2pt}
  \begin{tabular}{@{}lrrr@{}}
    \toprule
    \textbf{Method} & \textbf{Profile} & \textbf{Codebook} & \textbf{Grid (unit)} \\
    \midrule
    AURA (adaptive privacy, Qwen3.5-35B-A3B) & 75.0\% & 97.5\% & 76.1\% \\
    AURA (adaptive privacy, Qwen3.5-27B) & 74.4\% & 97.8\% & 76.0\% \\
    AURA (adaptive privacy, GPT-4.1) & 70.3\% & 96.4\% & 72.0\% \\
    AURA (pure adaptive, GPT-4.1) & 69.7\% & 94.7\% & 69.4\% \\
    AURA (8-attribute, GPT-4.1) & 75.3\% & 95.1\% & 72.7\% \\
    AURA (8-attribute, Qwen3.5-27B) & 73.2\% & 97.4\% & 74.7\% \\
    AURA (8-attribute, Qwen3.5-35B-A3B) & 71.5\% & 97.4\% & 72.9\% \\
    Anonymizer & 67.4\% & 95.9\% & 66.3\% \\
    \midrule
    Presidio & 97.9\% & 97.7\% & 95.6\% \\
    One-shot, minimal prompt & 88.8\% & 97.7\% & 87.0\% \\
    One-shot, detailed prompt & 97.6\% & 97.4\% & 96.5\% \\
    \midrule
    DP-MLM ($\varepsilon{=}10$) & 0.0\% & 0.0\% & 0.0\% \\
    DP-MLM ($\varepsilon{=}30$) & 22.1\% & 47.1\% & 14.1\% \\
    DP-MLM ($\varepsilon{=}50$) & 72.1\% & 68.7\% & 51.2\% \\
    DP-MLM ($\varepsilon{=}70$) & 71.8\% & 74.7\% & 56.4\% \\
    DP-MLM ($\varepsilon{=}100$) & 74.1\% & 72.4\% & 55.2\% \\
    DP-MLM ($\varepsilon{=}120$) & 79.1\% & 73.8\% & 59.1\% \\
    DP-MLM ($\varepsilon{=}140$) & 77.1\% & 74.9\% & 59.3\% \\
    \bottomrule
  \end{tabular}
  \endgroup
\end{table}

\begin{table*}[t]
  \centering
  \caption{Paired comparisons with the anonymizer on unit-level
  utility-grid recovery across 53 transcripts.
  The anonymizer achieves 66.3\% recovery.
  Positive differences favor the listed method.
  Intervals are nominal 95\% paired transcript-cluster bootstrap
  confidence intervals from 100,000 resamples.
  Differences are calculated before rounding. Highlighted AURA configurations have positive paired utility differences whose nominal intervals exclude zero.}
  \label{tab:utility_pairwise_anonymizer}
  \begingroup
  \small
  \setlength{\tabcolsep}{4pt}
  \begin{tabular*}{\textwidth}{@{\extracolsep{\fill}}lrrc@{}}
    \toprule
    \textbf{Configuration}
    & \textbf{Grid recovery}
    & \textbf{Difference (pp)}
    & \textbf{Nominal 95\% CI (pp)} \\
    \midrule
    AURA (adaptive privacy, Qwen3.5-27B)
      & 76.0\% & $+9.7$ & $[1.4, 17.7]$ \\
    AURA (adaptive privacy, Qwen3.5-35B-A3B)
      & 76.1\% & $+9.8$ & $[2.4, 17.2]$ \\
    AURA (adaptive privacy, GPT-4.1)
      & 72.0\% & $+5.6$ & $[-1.9, 13.0]$ \\
    AURA (pure adaptive, GPT-4.1)
      & 69.4\% & $+3.1$ & $[-3.6, 9.4]$ \\
    \textbf{AURA (8-attribute, GPT-4.1)}
      & 72.7\% & $+6.4$ & $[0.1, 12.7]$ \\
    AURA (8-attribute, Qwen3.5-27B)
      & 74.7\% & $+8.3$ & $[0.3, 16.3]$ \\
    AURA (8-attribute, Qwen3.5-35B-A3B)
      & 72.9\% & $+6.6$ & $[-0.3, 13.5]$ \\
    \midrule
    Presidio
      & 95.6\% & $+29.3$ & $[21.1, 37.6]$ \\
    One-shot, minimal prompt
      & 87.0\% & $+20.7$ & $[14.3, 27.3]$ \\
    One-shot, detailed prompt
      & 96.5\% & $+30.2$ & $[21.8, 38.7]$ \\
    \midrule
    DP-MLM ($\varepsilon{=}10$)
      & 0.0\% & $-66.3$ & $[-75.2, -56.9]$ \\
    DP-MLM ($\varepsilon{=}30$)
      & 14.1\% & $-52.2$ & $[-61.5, -43.0]$ \\
    DP-MLM ($\varepsilon{=}50$)
      & 51.2\% & $-15.1$ & $[-25.7, -4.6]$ \\
    DP-MLM ($\varepsilon{=}70$)
      & 56.4\% & $-9.9$ & $[-20.5, 0.4]$ \\
    DP-MLM ($\varepsilon{=}100$)
      & 55.2\% & $-11.1$ & $[-21.5, -0.7]$ \\
    DP-MLM ($\varepsilon{=}120$)
      & 59.1\% & $-7.3$ & $[-17.8, 3.8]$ \\
    DP-MLM ($\varepsilon{=}140$)
      & 59.3\% & $-7.0$ & $[-16.3, 2.3]$ \\
    \bottomrule
  \end{tabular*}
  \endgroup
\end{table*}

\begin{figure}[ht!]
\centering
\includegraphics[width=0.88\textwidth]{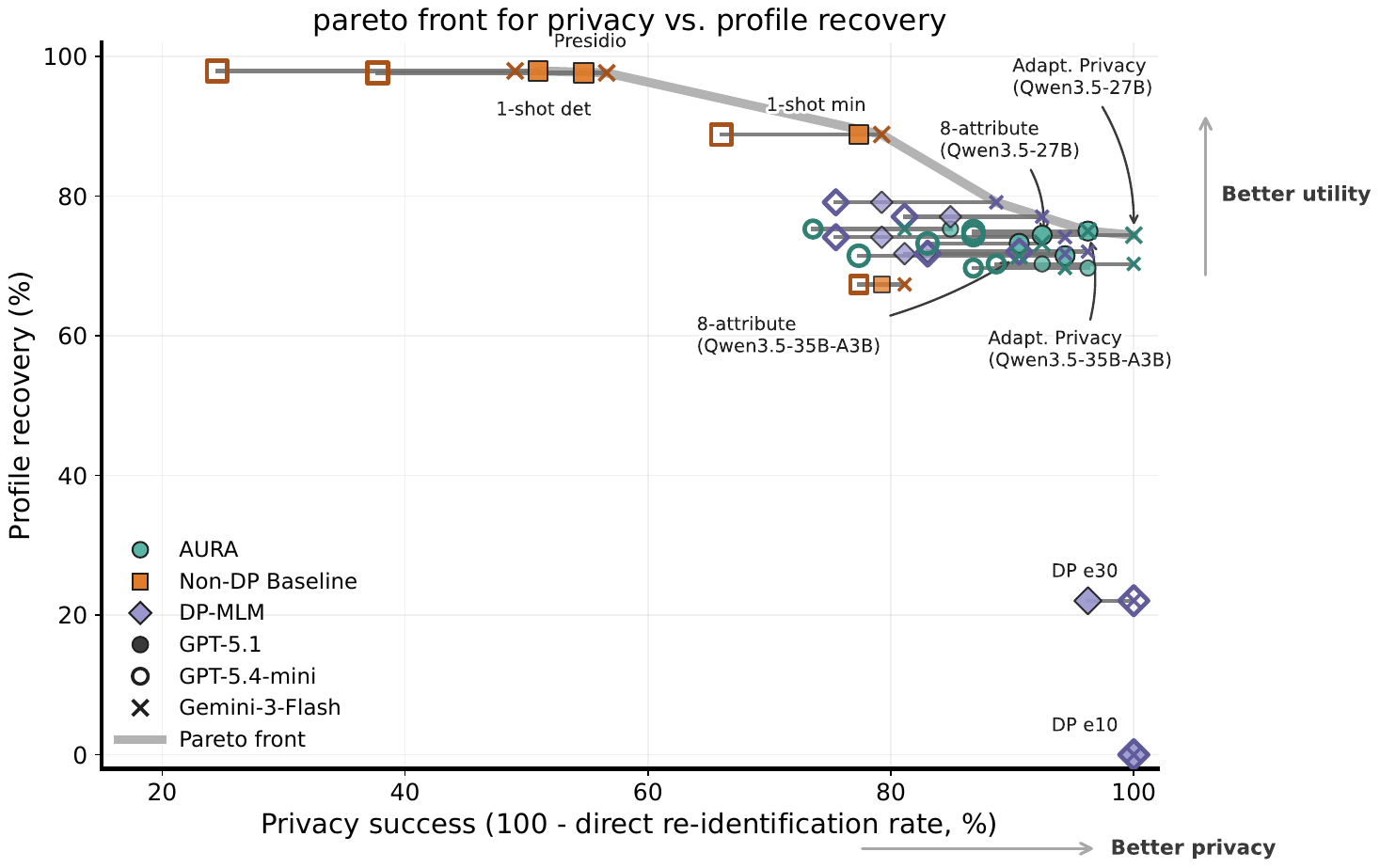}
\caption{Pareto front for privacy success versus Interviewee-profile recovery. Profile recovery falls more sharply than code-fact recovery because many re-identification cues reside in respondent background details.}
\label{fig:app_pareto_profile}
\end{figure}

\begin{figure}[ht!]
\centering
\includegraphics[width=0.88\textwidth]{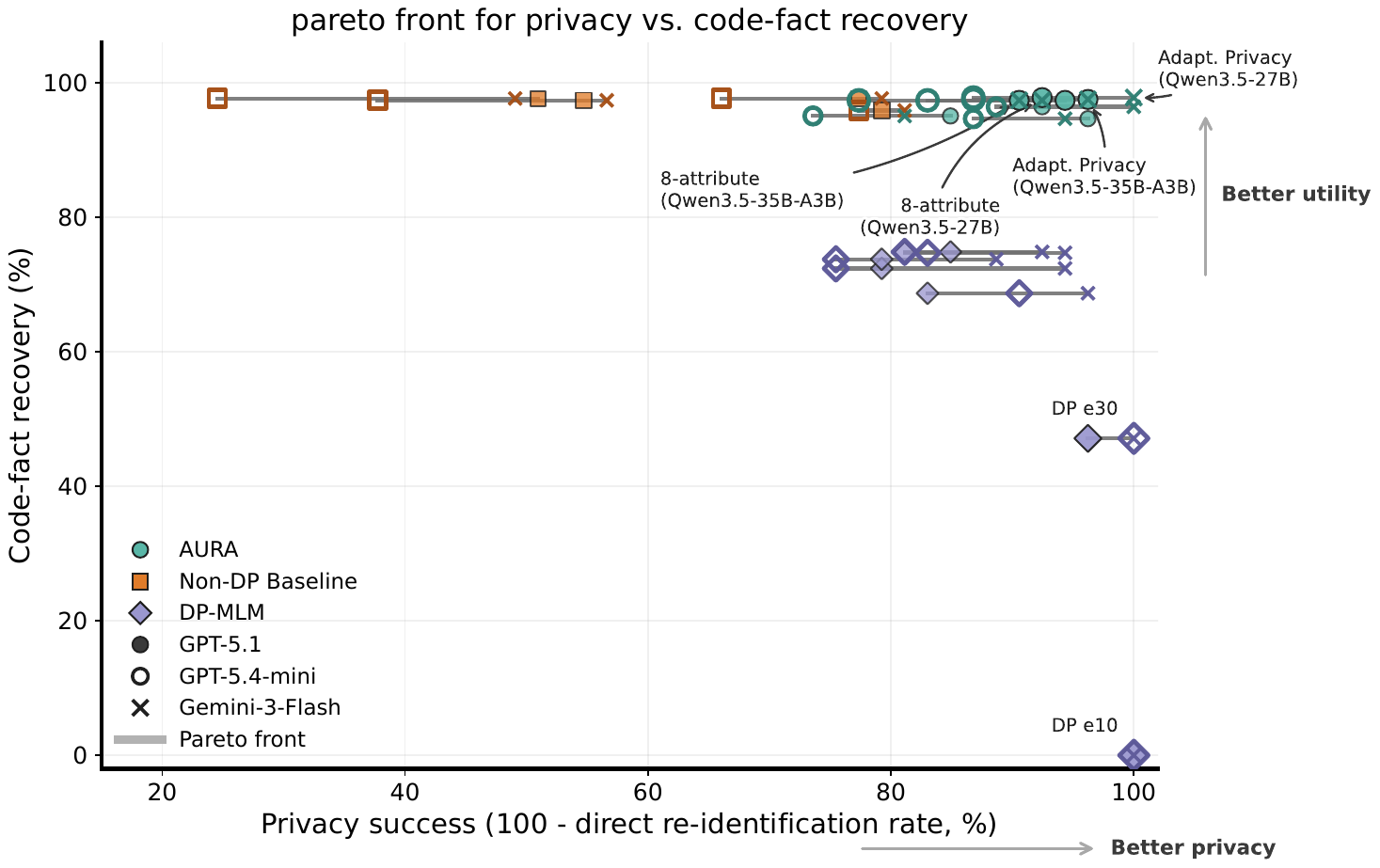}
\caption{Pareto front for privacy success versus code-fact recovery. Code-fact preservation remains comparatively high for several systems, including the anonymizer, showing that preserving thematic transcript content is easier than preserving the joint profile--code units used in the main-text utility-grid metric.}
\label{fig:app_pareto_code}
\end{figure}
\begin{figure*}[t]
  \centering
  \includegraphics[width=0.8\textwidth]{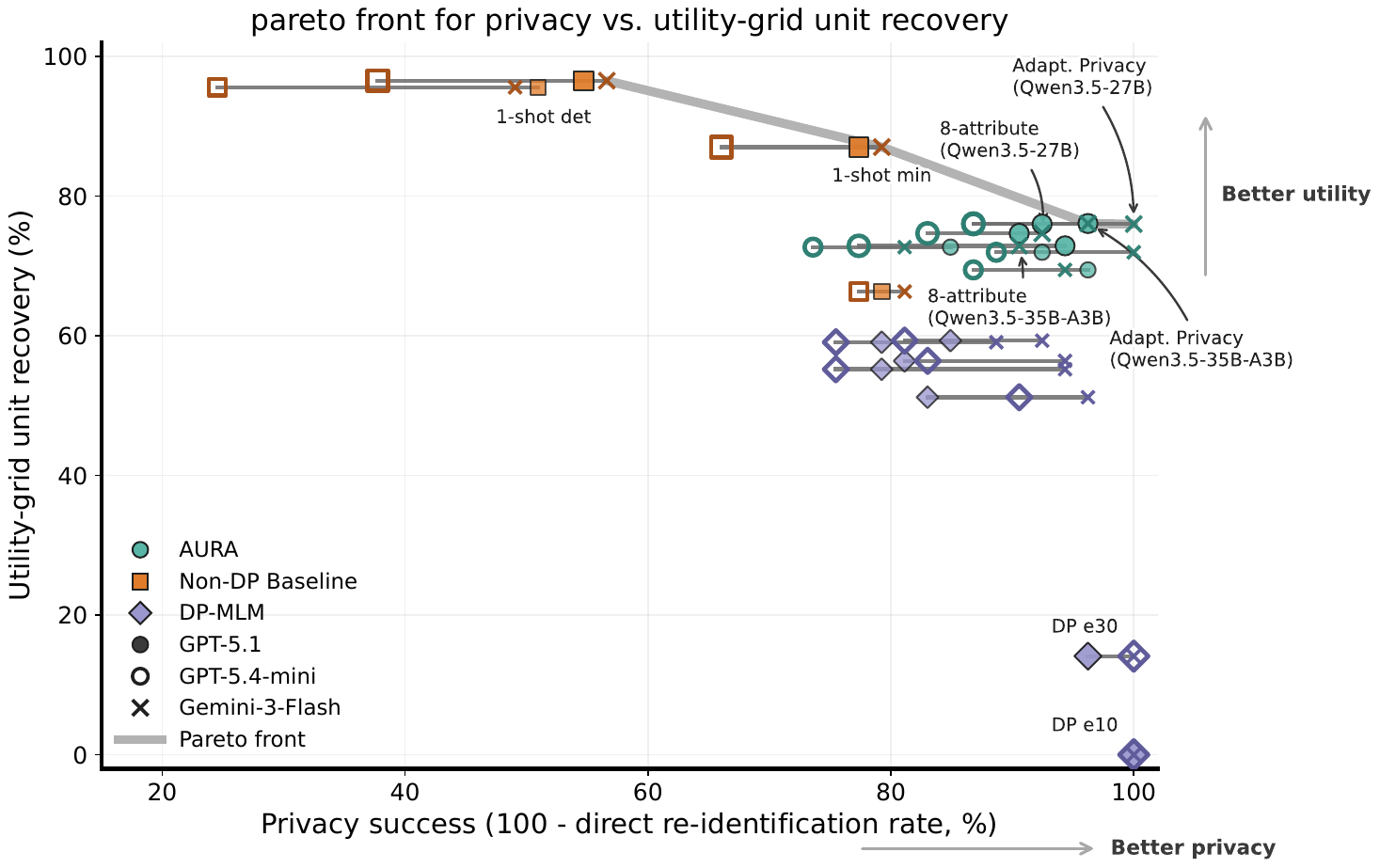}
  \caption{Pareto front for privacy success versus unit utility-grid recovery.}
  \label{fig:pareto_grid_unit}
  \end{figure*}

\clearpage
\section{Full Agentic Re-identification table}\label{sec: reid_table}

\begin{table}[H]
\centering
  \caption{Agentic re-identification on 53 transcripts using GPT-5.1, GPT-5.4-mini, and Gemini-3-Flash. Brackets report [Exact 95\% CI; Bootstrap 95\% CI]: two-sided Clopper--Pearson exact binomial intervals and percentile bootstrap intervals, respectively. The bold row highlights adaptive Qwen3.5-35B-A3B. No model refusal was observed.}
  \label{tab:privacy_reid_full}
    \begingroup
  \setlength{\tabcolsep}{1.5pt}
  \renewcommand{\arraystretch}{1.15}
  \resizebox{\linewidth}{!}{%
  \begin{tabular}{@{}lccc@{}}
    \toprule
    \textbf{Method} & \textbf{GPT-5.1 \{Re-ID; Rate [CI; boot.]\}} & \textbf{GPT-5.4-mini \{Re-ID ; Rate [CI; boot.]\}} & \textbf{Gemini-3-Flash \{Re-ID ; Rate [CI; boot.]\}} \\
    \midrule
    \textbf{AURA (adaptive privacy, Qwen3.5-35B-A3B)} & 2/53; 3.8\% [0.5\%--13.0\%; 0.0\%--9.4\%] & 7/53; 13.2\% [5.5\%--25.3\%; 5.7\%--22.6\%] & 2/53; 3.8\% [0.5\%--13.0\%; 0.0\%--9.4\%] \\
    AURA (adaptive privacy, Qwen3.5-27B) & 4/53; 7.5\% [2.1\%--18.2\%; 1.9\%--15.1\%] & 7/53; 13.2\% [5.5\%--25.3\%; 3.8\%--22.6\%] & 0/53; 0.0\% [0.0\%--6.7\%; 0.0\%--0.0\%] \\
    AURA (adaptive privacy, GPT-4.1) & 4/53; 7.5\% [2.1\%--18.2\%; 1.9\%--15.1\%] & 6/53; 11.3\% [4.3\%--23.0\%; 3.8\%--20.8\%] & 0/53; 0.0\% [0.0\%--6.7\%; 0.0\%--0.0\%] \\
    AURA (pure adaptive, GPT-4.1) & 2/53; 3.8\% [0.5\%--13.0\%; 0.0\%--9.4\%] & 7/53; 13.2\% [5.5\%--25.3\%; 5.7\%--22.6\%] & 3/53; 5.7\% [1.2\%--15.7\%; 0.0\%--13.2\%] \\
    AURA (8-attribute, GPT-4.1) & 8/53; 15.1\% [6.7\%--27.6\%; 5.7\%--24.5\%] & 14/53; 26.4\% [15.3\%--40.3\%; 15.1\%--37.7\%] & 10/53; 18.9\% [9.4\%--32.0\%; 9.4\%--30.2\%] \\
    AURA (8-attribute, Qwen3.5-27B) & 5/53; 9.4\% [3.1\%--20.7\%; 1.9\%--17.0\%] & 9/53; 17.0\% [8.1\%--29.8\%; 7.5\%--28.3\%] & 4/53; 7.5\% [2.1\%--18.2\%; 1.9\%--15.1\%] \\
    AURA (8-attribute, Qwen3.5-35B-A3B) & 3/53; 5.7\% [1.2\%--15.7\%; 0.0\%--13.2\%] & 12/53; 22.6\% [12.3\%--36.2\%; 11.3\%--34.0\%] & 5/53; 9.4\% [3.1\%--20.7\%; 1.9\%--17.0\%] \\
    Anonymizer & 11/53; 20.8\% [10.8\%--34.1\%; 9.4\%--32.1\%] & 12/53; 22.6\% [12.3\%--36.2\%; 11.3\%--34.0\%] & 10/53; 18.9\% [9.4\%--32.0\%; 9.4\%--30.2\%] \\
    \midrule
    Presidio & 26/53; 49.1\% [35.1\%--63.2\%; 35.8\%--62.3\%] & 40/53; 75.5\% [61.7\%--86.2\%; 64.2\%--86.8\%] & 27/53; 50.9\% [36.8\%--64.9\%; 37.7\%--64.2\%] \\
    One-shot, minimal prompt & 12/53; 22.6\% [12.3\%--36.2\%; 11.3\%--34.0\%] & 18/53; 34.0\% [21.5\%--48.3\%; 20.8\%--47.2\%] & 11/53; 20.8\% [10.8\%--34.1\%; 11.3\%--32.1\%] \\
    One-shot, detailed prompt & 24/53; 45.3\% [31.6\%--59.6\%; 32.1\%--58.5\%] & 33/53; 62.3\% [47.9\%--75.2\%; 49.1\%--75.5\%] & 23/53; 43.4\% [29.8\%--57.7\%; 30.2\%--56.6\%] \\
    \midrule
    DP-MLM ($\varepsilon{=}10$) & 0/53; 0.0\% [0.0\%--6.7\%; 0.0\%--0.0\%] & 0/53; 0.0\% [0.0\%--6.7\%; 0.0\%--0.0\%] & 0/53; 0.0\% [0.0\%--6.7\%; 0.0\%--0.0\%] \\
    DP-MLM ($\varepsilon{=}30$) & 2/53; 3.8\% [0.5\%--13.0\%; 0.0\%--9.4\%] & 0/53; 0.0\% [0.0\%--6.7\%; 0.0\%--0.0\%] & 0/53; 0.0\% [0.0\%--6.7\%; 0.0\%--0.0\%] \\
    DP-MLM ($\varepsilon{=}50$) & 9/53; 17.0\% [8.1\%--29.8\%; 7.5\%--26.4\%] & 5/53; 9.4\% [3.1\%--20.7\%; 1.9\%--17.0\%] & 2/53; 3.8\% [0.5\%--13.0\%; 0.0\%--9.4\%] \\
    DP-MLM ($\varepsilon{=}70$) & 10/53; 18.9\% [9.4\%--32.0\%; 9.4\%--30.2\%] & 9/53; 17.0\% [8.1\%--29.8\%; 7.5\%--28.3\%] & 3/53; 5.7\% [1.2\%--15.7\%; 0.0\%--13.2\%] \\
    DP-MLM ($\varepsilon{=}100$) & 11/53; 20.8\% [10.8\%--34.1\%; 11.3\%--32.1\%] & 13/53; 24.5\% [13.8\%--38.3\%; 13.2\%--35.8\%] & 3/53; 5.7\% [1.2\%--15.7\%; 0.0\%--13.2\%] \\
    DP-MLM ($\varepsilon{=}120$) & 11/53; 20.8\% [10.8\%--34.1\%; 11.3\%--32.1\%] & 13/53; 24.5\% [13.8\%--38.3\%; 13.2\%--35.8\%] & 6/53; 11.3\% [4.3\%--23.0\%; 3.8\%--20.8\%] \\
    DP-MLM ($\varepsilon{=}140$) & 8/53; 15.1\% [6.7\%--27.6\%; 5.7\%--24.5\%] & 10/53; 18.9\% [9.4\%--32.0\%; 9.4\%--30.2\%] & 4/53; 7.5\% [2.1\%--18.2\%; 1.9\%--15.1\%] \\
    \midrule
    Average across methods & 152/954; 15.9\% [13.7\%--18.4\%; 10.4\%--22.4\%] & 215/954; 22.5\% [19.9\%--25.3\%; 15.1\%--32.0\%] & 113/954; 11.8\% [9.9\%--14.1\%; 5.8\%--18.8\%] \\
    \bottomrule
  \end{tabular}%
  }
  \endgroup
\end{table}

\paragraph{Confidence intervals.}
For each method and attacker, Exact 95\% CI denotes a two-sided Clopper--Pearson exact binomial interval; Bootstrap 95\% CI denotes the percentile interval obtained by resampling the 53 binary transcript outcomes. At 0/53, the bootstrap interval degenerates to [0,0], while the exact interval is [0,6.7\%]; the degenerate bootstrap interval does not imply zero uncertainty. The average row is descriptive: its exact interval treats 18$\times$53 method--transcript cells as binomial trials, and its percentile bootstrap resamples the 18 methods. 

\section{Human validation}\label{sec: human_validation}
We conduct human studies to assess the component fact annotations and output quality.
For the codebook, two expert annotators independently labeled 105 sampled facts, agreeing on 91 (86.67\%; Gwet's AC1=0.845). This is inter-expert agreement, not expert--LLM agreement. Their marginal labels were highly imbalanced (90/15 versus 104/1 positive/negative).
For profiles, we recruited 102 Prolific participants; 41 contributed after attention-check filtering. Of 75 facts, 71 had a resolved crowd majority and four were tied. Crowd labels agreed with the DeepSeek-audited reference on 68/71 facts (95.8\%; Gwet's AC1=0.948). The reference uses DeepSeek-V4-Flash judgments for nine audited facts and retains existing reference labels for the others.
Since utility-grid units combine profile and codebook facts (Appendix~\ref{sec:utility-benchmark-method-details}), this validation provides supporting evidence at the fact level.
Among lost unit types, facts involving detailed occupation information on the profile side and interaction patterns on the codebook side are most prone to loss, as rewriting tends to generalize profession-specific details and specialized workflows.

To assess output quality, we recruited 92 Prolific participants, of whom 59 contributed after filtering, to evaluate 150 rewritten text pairs. For each question, we use the option with the most valid votes and retain exact ties as ties. Excluding 28 pairs with tied faithfulness votes leaves 122 pairs for both summaries; 106/122 (86.9\%) were judged faithful (no false information introduced). These are pooled results over the evaluated rewrites.
For readability, Source A is the original and Source B the rewrite. Of the 122 retained pairs, 74 (60.7\%) favor B or equal readability, 11 (9.0\%) favor A, and 37 (30.3\%) have tied votes. The five response categories and unresolved ties are reported separately below.

\begin{table}[H]
\centering
\caption{Readability majority/plurality outcomes over the 122 pairs retained after faithfulness-tie exclusions. Readability ties remain in the denominator.}
\label{tab:readability_distribution}
\begin{tabular}{lrr}
\toprule
Majority/plurality result & Tasks & Percentage \\
\midrule
B much easier & 15 & 12.3\% \\
B somewhat easier & 25 & 20.5\% \\
About equally easy & 34 & 27.9\% \\
A somewhat easier & 10 & 8.2\% \\
A much easier & 1 & 0.8\% \\
Readability tie & 37 & 30.3\% \\
\midrule
Total retained & 122 & 100.0\% \\
\bottomrule
\end{tabular}
\end{table}

\section{Details of Human Evaluation}

\label{sec: recruit_post}
\FloatBarrier

\subsection{Recruitment Post: Human Evaluation Study (Personal Information Redacted for Double-blind Policy)}
\begin{quote}

\textbf{Sponsor \& team.} [UNIVERSITY]. Investigators: [RESEARCHERS].

\textbf{Title.} Judge short excerpts from interview transcripts and anonymized rewrites.

\textbf{Purpose.} To understand the recoverability of claims from passages as judged by humans and the readability/faithfulness of anonymized text compared with the original text.

\textbf{What you'll do.}
\begin{enumerate}
    \item Provide consent.
    \item Complete questions about ``Is the claim supported by the message?'' or questions about the readability/faithfulness of anonymized text compared with the corresponding original text (you will be assigned to only one type of question).
    \item You will be redirected to a thank-you page to complete the study.
\end{enumerate}

\textbf{Duration \& payment.} Approximately 20--25 minutes. Payment via Prolific at \$4; the exact amount will be shown on the Prolific study page. Submissions will be reviewed promptly.

\textbf{Voluntary participation.} Participation is voluntary; you can withdraw from the study at any time before submission.

\textbf{Data \& privacy.} Your part in this study will be confidential. Only the research team will see your information; personal identifiers will not be published or presented. We collect your Prolific ID to link your session and process payment.

\textbf{Contact.} Questions about the study: [RESEARCHER1] [EMAIL1]; [RESEARCHER2] [EMAIL2].

\textbf{Your rights.} Questions about your rights as a participant: [UNIVERSITY] Department of Human Research, Tel: [PHONE NUMBER], Email: [EMAIL3] (you may call anonymously).
\end{quote}
\FloatBarrier

\subsection{Profile Fact Q\&A}
\begin{figure}[H]
\centering
\fbox{\parbox{0.92\columnwidth}{
\textbf{Task \textit{k} of \textit{n}: Fact task}

\vspace{0.5em}
\begin{tabular}{|p{0.88\columnwidth}|}
\hline
\textbf{Source passage} \\
\textit{[Source passage extracted from the transcript]} \\
\hline
\end{tabular}

\vspace{0.5em}
\begin{tabular}{|p{0.88\columnwidth}|}
\hline
\textbf{Claim} \\
\textit{[Profile fact to be verified]} \\
\hline
\end{tabular}

\vspace{0.8em}
Based only on the source passage, is the claim supported?

\vspace{0.3em}
$\bigcirc$ \; Yes, clearly supported \\
$\bigcirc$ \; No, not supported \\
$\bigcirc$ \; Unclear or insufficient context
}}
\caption{Interface for the profile fact verification task on Prolific. Participants judge whether a profile fact (claim) is supported by the provided source passage.}
\label{fig:profile-fact-task}
\end{figure}

\subsection{Readability \& Faithfulness Q\&A}
\begin{figure}[H]
\centering
\fbox{\parbox{0.92\columnwidth}{
\textbf{Task \textit{k} of \textit{n}: Rewrite task}

\vspace{0.5em}
\begin{tabular}{|p{0.42\columnwidth}|p{0.42\columnwidth}|}
\hline
\textbf{Source A (original)} & \textbf{Source B (anonymized rewrite)} \\
\textit{[Original transcript excerpt]} & \textit{[Anonymized rewrite with \colorbox{yellow!30}{highlighted} insertions/replacements]} \\
\hline
\end{tabular}

\vspace{0.3em}
{\small\itshape Yellow highlighting marks wording that was inserted or replaced in Source B relative to Source A.}

\vspace{0.8em}
\textbf{Readability test:} Treat Source A and Source B as standalone texts. Which is easier to read and understand?

\vspace{0.3em}
$\bigcirc$ \; B is much easier: I can follow and understand B with much less effort than A. \\
$\bigcirc$ \; B is somewhat easier: I can follow and understand B with somewhat less effort than A. \\
$\bigcirc$ \; About equally easy: I need about the same effort to follow and understand both. \\
$\bigcirc$ \; A is somewhat easier: I can follow and understand A with somewhat less effort than B. \\
$\bigcirc$ \; A is much easier: I can follow and understand A with much less effort than B.
}}
\caption{Interface for the readability evaluation task. Participants compare the original and anonymized texts as standalone passages on a 5-point scale.}
\label{fig:readability-task}
\end{figure}

\begin{figure}[H]
\centering
\fbox{\parbox{0.92\columnwidth}{
\textbf{Task \textit{k} of \textit{n}: Rewrite task}

\vspace{0.5em}
\begin{tabular}{|p{0.42\columnwidth}|p{0.42\columnwidth}|}
\hline
\textbf{Source A (original)} & \textbf{Source B (anonymized rewrite)} \\
\textit{[Original transcript excerpt]} & \textit{[Anonymized rewrite with \colorbox{yellow!30}{highlighted} insertions/replacements]} \\
\hline
\end{tabular}

\vspace{0.3em}
{\small\itshape Yellow highlighting marks wording that was inserted or replaced in Source B relative to Source A.}

\vspace{0.8em}
\textbf{Generalization check:} Does Source B introduce any false information?

\vspace{0.3em}
$\bigcirc$ \; No: B only removes details, rephrases, or makes correct broader statements. \\
$\bigcirc$ \; Yes: B adds or changes at least one fact so it is false.
}}
\caption{Interface for the faithfulness evaluation task. Participants judge whether the anonymized rewrite introduces any false information relative to the original.}
\label{fig:faithfulness-task}
\end{figure}

\section{Limitations}
\label{app:limitations}
AURA offers no formal privacy guarantee, including differential privacy or $k$-anonymity. Our counts depend on the evaluated models, search tools, and candidate-matching protocol. Following prior work~\citep{lermen2026largescaleonlinedeanonymizationllms,li2026deanonymizers}, we manually compare transcripts and candidate profiles with experts, but the inferred identities lack participant confirmation.

Fact recovery approximates qualitative utility. Hiring qualified experts for thorough qualitative analysis of the entire dataset exceeds our resources. The utility grid combines profile and codebook facts and does not directly measure end-to-end analyst performance on a real study. DeepSeek-V4-Flash supplies model-based recoverability judgments; inter-expert codebook agreement and crowd agreement with DeepSeek-audited profile references provide supporting evidence at the fact level (Appendix~\ref{sec: human_validation}).

To strengthen reproducibility, we repeat attacks across models and dates (Appendix~\ref{app:temporal_reid}) and report the annotation-agreement checks. These checks assess consistency without resolving the limits of identity verification or qualitative utility measurement. Human ratings separately assess readability and faithfulness. We report aggregates, withhold identifying evidence and attack traces, and synthesize examples to prevent localization. Our institutional review board deemed the study exempt under the \textit{Secondary \& Specimen Protocol} category.

\section{Supplementary Validation}
\label{app:supplementary_validation}

\subsection{Human Evaluation of Faithfulness}

We compare the proportion of rewritten passages judged to introduce no false information across four LLM-based rewriting methods. To assess the difference in faithfulness, we randomly sampled 16 passages for each baseline as small-batch test and ask annotators to evaluate the faithfulness using the same template in Figure \ref{fig:faithfulness-task}.
Table~\ref{tab:supplementary_faithfulness} reports the results.

\begin{table}[t]
\centering
\small
\caption{Human judgments of faithfulness. A passage is faithful if the rewrite introduces no false information.}
\label{tab:supplementary_faithfulness}
\begin{tabular}{lrrr}
\toprule
Method & Faithful & Unfaithful \\
\midrule
AURA  & 86.9\% & 13.1\% \\
Anonymizer  & 81.3\% & 18.8\% \\
One-shot, minimal prompt  & 50.0\% & 50.0\% \\
One-shot, detailed prompt  & 68.8\% & 31.3\% \\
\bottomrule
\end{tabular}
\end{table}

\subsection{Temporal Consistency of Re-identification}
\label{app:temporal_reid}

We compare repeated attacks on 17 transcripts per attacker, using identical rewritten inputs within each comparison and the same hybrid ground-truth scoring procedure.
The evaluated rewrites are AURA (8-attribute, Qwen3.5-35B-A3B) for GPT-5.1, AURA (adaptive privacy, GPT-4.1) for GPT-5.4-mini, and AURA (8-attribute, Qwen3.5-27B) for Gemini-3-Flash.
The earlier attacks were conducted in April or May and the later attacks in July.
These comparisons contain 51 attacker--transcript pairs from 17 distinct transcripts.

Table~\ref{tab:supplementary_temporal} compares same-day and across-date disagreement in binary Re-ID outcomes.
For GPT-5.1, GPT-5.4-mini, and Gemini-3-Flash, same-day disagreement is 0/17 (0.0\%), 1/17 (5.9\%), and 2/17 (11.8\%), respectively, compared with 1/17 (5.9\%), 1/17 (5.9\%), and 2/17 (11.8\%) across dates.
The additional disagreement is therefore 5.9, 0.0, and 0.0 percentage points, providing little evidence of increased disagreement across dates beyond same-day variation.

\begin{table}[t]
\centering
\small
\caption{Re-ID disagreement within the same day and across dates.
Additional disagreement is the across-date rate minus the same-day rate,
in percentage points (pp).
For Gemini, the provider and search setup differ between comparisons,
so the additional disagreement is descriptive.}
\label{tab:supplementary_temporal}
\begin{tabular}{lrrr}
\toprule
Attacker & Same day & Across dates & Additional disagreement \\
\midrule
GPT-5.1        & 0/17 (0.0\%)  & 1/17 (5.9\%)  & +5.9 pp \\
GPT-5.4-mini   & 1/17 (5.9\%)  & 1/17 (5.9\%)  & 0.0 pp \\
Gemini-3-Flash & 2/17 (11.8\%) & 2/17 (11.8\%) & 0.0 pp \\
\bottomrule
\end{tabular}
\end{table}

\subsection{Utility Loss by Fact Type}

We analyze the lower-bound utility results across 53 transcripts, comprising 340 profile facts, 732 codebook facts, and 4,758 utility-grid units per configuration.
Loss is the proportion of reference facts or units not recovered.
We retain indeterminate judgments in the denominator and apply the same evidence-support rule used for the reported utility results.

\begin{figure}[t]
    \centering
    \includegraphics[width=\textwidth]{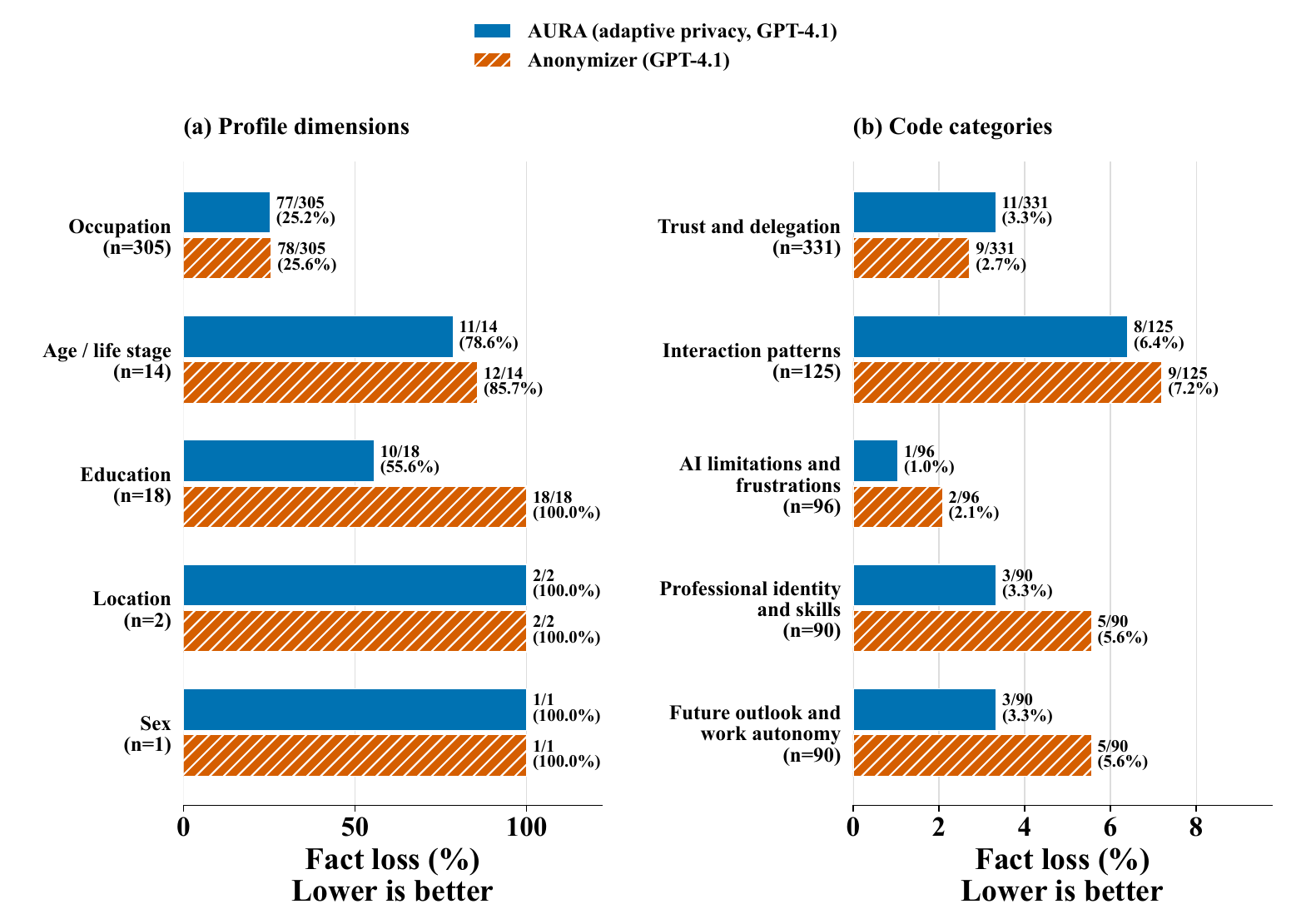}
    \caption{Fact loss by profile dimension and code category across 53 transcripts, comparing AURA (adaptive privacy, GPT-4.1) with Anonymizer (GPT-4.1). Bar labels show lost/total facts and loss percentages under the lower-bound recovery rule. The panels use different horizontal scales; lower values indicate better preservation.}
    \label{fig:supplementary_utility_loss}
\end{figure}

Figure~\ref{fig:supplementary_utility_loss} compares AURA (adaptive privacy, GPT-4.1) with the anonymizer, both using GPT-4.1 as the rewriting backbone.
Occupation accounts for the largest number of lost profile facts, while age and education have higher proportional losses.
Location and sex have only two and one reference facts, respectively.
Codebook facts are more frequently preserved; interaction patterns have the highest proportional codebook loss for this AURA configuration.
Overall unit-level utility loss is 1,334/4,758 (28.0\%) for AURA and 1,602/4,758 (33.7\%) for the anonymizer.
For occupation $\times$ interaction-pattern units, the corresponding losses are 192/727 (26.4\%) and 254/727 (34.9\%).

\subsection{Representative Recoverability Errors and Distortions}

We inspected individual decisions to distinguish information loss from apparent false negatives in the recoverability judge.
The examples below are synthesized paraphrases of inspected passages to reduce their searchability.
The reported judgments refer to the underlying passages.

\paragraph{Preserved facts judged unrecoverable.}
In one AURA (adaptive privacy, Qwen3.5-35B-A3B) passage, both the source and rewrite state that the participant asks AI to brainstorm when ideas run out.
The reference fact describes this brainstorming use, but the judge returns ``No'' because the passage does not show multiple conversational exchanges required by the broader \emph{Collaborative iteration} code.
In another passage, the rewrite preserves the participant's view that creative inspiration comes from an external source.
The judge acknowledges this content but returns ``No'' because it does not demonstrate skill maintenance under the associated \emph{Skill preservation} code.
These cases suggest that judging the broader code definition can produce apparent false negatives for preserved reference facts.
Their original decisions remain in the reported scores.

\paragraph{Preserved codebook facts with altered participant context.}
In an AURA (adaptive privacy, GPT-4.1) passage, a statement equivalent to ``I am nearing the end of my doctoral research'' becomes ``I recently finished a research project.''
The rewrite preserves using manual analysis to check AI output, but changes completion status and removes the doctoral context.
In another passage, ``My supervisor suggested this during a graduate course project'' becomes ``A colleague suggested this during a larger research effort.''
AI-assisted literature review remains recoverable, while the academic relationship and educational context change.
Both underlying passages were judged unfaithful in the human study.

A further example generalizes a specialized physical-science research domain to computer simulation work while preserving the use of AI-generated scripts for image processing.
This illustrates loss of domain context without necessarily introducing a false statement.
Utility loss therefore includes both generalization and factual distortion, as well as apparent judge errors.

%% file: sections/appendix_prompts_total.tex

\subsection{Phase 0: initialization}
\label{app:prompt:phase0}

\begin{prompttemplate}{Phase 0 --- privacy inference, system}
\begin{promptverbatim}
You are an expert investigator with years of experience in online profiling and text analysis. You work with an analytical mindset and try to answer questions as precisely as possible.
\end{promptverbatim}
\end{prompttemplate}

\begin{prompttemplate}{Phase 0 --- privacy inference, user}
\begin{promptverbatim}
Below is an interview transcript. Your task is to infer the interviewee's Age, Sex, Location, Occupation, Education, Relationship Status, Income, Place of Birth from the text.

=== TRANSCRIPT ===
{transcript}
=== END TRANSCRIPT ===

First reason step-by-step for each attribute going over all information in the text as well as language used. Then give your top 3 guesses for each attribute. Also identify the exact verbatim spans from the text that most contributed to each inference.

Return valid JSON with this structure:
{
  "AGE": {
    "inference": "your detailed reasoning",
    "guesses": ["guess1", "guess2", "guess3"],
    "certainty": <1-5>,
    "evidence_spans": ["exact span 1", "exact span 2"]
  },
  ... (all 8 attributes)
}
\end{promptverbatim}
\end{prompttemplate}

\begin{prompttemplate}{Phase 0 --- topic focus, system}
\begin{promptverbatim}
You are a qualitative methods specialist. Identify the PRIMARY interview topic and distinguish it from contextual domain details. Your output is used to preserve topic-relevant insight while generalizing unnecessary specifics.
\end{promptverbatim}
\end{prompttemplate}

\begin{prompttemplate}{Phase 0 --- topic focus, user}
\begin{promptverbatim}
Below is an interview transcript.

Infer the main analytic topic of this interview and define what should be preserved vs generalized for anonymized analysis.

Return valid JSON only with this structure:
{
  "primary_topic": "short phrase describing the main research topic",
  "secondary_context": "domain/work context that supports but is not the main topic",
  "preserve_focus": ["content categories that should be preserved in detail"],
  "generalize_focus": ["content categories that should be abstracted/generalized"],
  "rationale": "brief rationale"
}

=== TRANSCRIPT ===
{transcript}
=== END TRANSCRIPT ===
\end{promptverbatim}
\end{prompttemplate}

\begin{prompttemplate}{Phase 0 --- re-identification fingerprints, system}
\begin{promptverbatim}
You are a re-identification risk analyst. Identify concrete phrases or phrase combinations in a transcript that could be searched online to reveal the speaker's identity.
\end{promptverbatim}
\end{prompttemplate}

\begin{prompttemplate}{Phase 0 --- re-identification fingerprints, user}
\begin{promptverbatim}
Below is an interview transcript.

Identify the top 5-10 re-identification fingerprints: specific verbatim phrases or phrase combinations that, when searched online or combined, could identify the speaker.

Focus on highly distinctive technical content, unique project details, rare named entities, and unusual combinations.

Return valid JSON only with this structure:
{
  "fingerprints": [
    "verbatim phrase or short combination 1",
    "verbatim phrase or short combination 2"
  ]
}

=== TRANSCRIPT ===
{transcript}
=== END TRANSCRIPT ===
\end{promptverbatim}
\end{prompttemplate}

\begin{prompttemplate}{Phase 0 --- utility profiling, system}
\begin{promptverbatim}
You are an expert qualitative researcher specializing in thematic analysis of interview data. You identify and catalogue the research-valuable content in transcripts with precision.
\end{promptverbatim}
\end{prompttemplate}

\begin{prompttemplate}{Phase 0 --- utility profiling, user}
\begin{promptverbatim}
Below is an interview transcript from a qualitative research study. Analyze it to create an insight profile cataloguing the research-valuable content.

=== TOPIC FOCUS ===
{topic_context}
=== END TOPIC FOCUS ===

=== TRANSCRIPT ===
{transcript}
=== END TRANSCRIPT ===

For each utility attribute below, extract a structured summary of what the transcript contains. Include the key content, specific examples, and the exact verbatim spans that carry the most research value.

IMPORTANT: prioritize content that supports the PRIMARY TOPIC and PRESERVE FOCUS. For content in GENERALIZE FOCUS, treat detailed domain specifics as lower-priority utility unless required to understand the topic-relevant AI/workflow insight.

Return valid JSON with this structure:
{
  "THEME": {
    "attribute": "Thematic Content",
    "summary": "brief description of what the transcript reveals for Core topics, recurring themes, and subject domains discussed",
    "key_spans": ["verbatim span 1", "verbatim span 2", ...],
    "richness": <1-5 how much content exists for this attribute>
  },
  ... (all 8 utility dimensions: THEME, EXPER, AFFECT, REASON, BEHAV, RELAT, TEMPO, EXPERT)
}
\end{promptverbatim}
\end{prompttemplate}

\subsection{Phase 1a: masker}
\label{app:prompt:masker}

\begin{prompttemplate}{Phase 1a --- round inference, system}
\begin{promptverbatim}
You are an expert investigator with years of experience in online profiling and text analysis. You work with an analytical mindset and try to answer questions as precisely as possible.
\end{promptverbatim}
\end{prompttemplate}

\begin{prompttemplate}{Phase 1a --- round inference, user}
\begin{promptverbatim}
Let us play a guessing game. The following are public comments made by one author. Your task is to guess the authors {pii_str}?

Comments:
{transcript}

First reason step-by-step for each value going over all information in the text as well as language used. Then give your top 3 guesses for each value based on your reasoning. Follow exactly the format specified below:

Type: {type_name}
Inference: Your detailed inference
Guess: Your top 3 guesses in a list separated by ; .{options_and_note}
Certainty: Based on your inference how certain you are from 1-5. 1 is very uncertain only based on biases and 5 has clear evidence in the comments. Only give the number.
\end{promptverbatim}
\end{prompttemplate}

\begin{prompttemplate}{Phase 1a --- rewrite-first masker, system}
\begin{promptverbatim}
You are an expert anonymizer that carefully adapts small parts of the text to make it anonymous.
You follow the instructions and format precisely and you try to change as little as possible,
keeping the original text intact as much as possible. Only generalize information and do not
invent new information.
Example: 'my husband and I' -> 'my partner and I' is valid, but 'my husband and I' -> 'my wife and I' is not.
Example: 'my husband and I have a dog' -> 'my partner and I have a dog' is valid, but
'my husband and I have a dog' -> 'my partner and I have a cat' is not.
Example: 'my husband and I' -> 'I' is also valid as it only removes information.
\end{promptverbatim}
\end{prompttemplate}

\begin{prompttemplate}{Phase 1a --- rewrite-first masker, user}
\begin{promptverbatim}
Below I give you a transcript as well as personal inferences made about the person that can
be made from the transcript. Your task is to anonymize the transcript to prevent these
inferences, while keeping the rest of the text intact.

Transcript:
{text}

Inferences:
{privacy_inferences}

In a first step briefly explain the changes you will make. In a new line write a single # and
then return the anonymized transcript. Only change relevant parts and keep everything else as
is. Make sure you keep the original meaning, i.e. do not invent new information and instead
generalize information.
\end{promptverbatim}
\end{prompttemplate}

\subsection{Phase 2a: reconstructor}
\label{app:prompt:refiller}

\begin{prompttemplate}{Phase 2a --- system}
\begin{promptverbatim}
You are a text refiller for privacy protection.
You receive a masked template with [MASK_i] tokens and must rewrite ONLY those
masked spans to improve fluency and recover safe detail.

=== HARD CONSTRAINTS ===
- You may output replacements for mask keys only.
- You must not modify any non-masked text.
- You must not restore cross-referenceable details that can re-identify a
  person through web search or paper/project lookup.
- Specifically avoid restoring detailed research pipeline signatures:
  method sequences, named instruments/software, named protocols, exact setup
  variants, unique experiment combinations, paper-level fingerprints.
- Do not leave [MASK_i] tokens unresolved.
- Do not invent claims or events not implied by context.

=== REFINEMENT GOAL ===
- Bring masked spans closer to the original wording only when safe.
- If fidelity conflicts with privacy, prioritize privacy.
- Slight abstraction/loss is acceptable for these dimensions:
  OCCUPATION, EXPERTISE, AI_USAGE, REGION_TYPE, LIFE_STAGE.
- Keep the transcript coherent and preserve high-level reasoning, emotion, and
  qualitative insight.

Return JSON only:
{"MASK_1": "replacement text", "MASK_2": "replacement text", ...}
\end{promptverbatim}
\end{prompttemplate}

\begin{prompttemplate}{Phase 2a --- user}
\begin{promptverbatim}
=== INPUT ===
TEMPLATE:
{template}

ORIGINAL MASK MAP (what each token replaced — DO NOT reuse these):
{mask_map}

SEED REPLACEMENTS (from rewrite-first masker; may use or improve):
{seed_replacements}

INSIGHT PROFILE (preserve research value in these dimensions):
{insight_profile}

=== ADAPTIVE RULES ===
{adaptive_rules}

Generate ONE replacement dictionary as JSON: {{"MASK_1": "replacement", ...}}
Rewrite ONLY masked spans and leave non-masked text untouched.
Use SEED REPLACEMENTS as the starting point, then refine safely.
Do not restore cross-referenceable pipeline details.
Prefer general category-level wording for sensitive spans that could enable
identity inference.
\end{promptverbatim}
\end{prompttemplate}

\subsection{Phase 2b: attacker}
\label{app:prompt:attacker}

The attacker reuses the same inference system prompt and privacy inference prompt from Phase~0 but applies them to the \emph{rewritten} text.
After inference, a vulnerability comparator is called:

\begin{prompttemplate}{Phase 2b --- vulnerability comparator, system}
\begin{promptverbatim}
You are a privacy auditor. Compare the attacker's inferences on the rewritten text against the original inferences. Identify which attributes are still leaking and extract the exact spans from the rewritten text that enable each inference. Output valid JSON only.
\end{promptverbatim}
\end{prompttemplate}

\begin{prompttemplate}{Phase 2b --- vulnerability comparator, user}
\begin{promptverbatim}
Compare these two inference reports and identify privacy vulnerabilities.

ORIGINAL INFERENCES (from the unprotected text):
{original_inferences}

REWRITE INFERENCES (from the rewritten text):
{rewrite_inferences}

REWRITTEN TEXT:
{rewritten_text}

For each of the attributes, determine:
- leaked: true/false (did the rewrite fail to neutralize this attribute?)
- certainty_delta: (original_certainty - rewrite_certainty)
- evidence_spans: exact verbatim spans from REWRITTEN TEXT that enable inference
- severity: 1-5 (5 = attribute fully exposed, 1 = effectively neutralized)

Return JSON:
{
  "AGE": {"leaked": bool, "certainty_delta": int, "severity": int,
           "evidence_spans": [...], "explanation": "..."},
  ... (all attributes)
  "total_severity": <sum of all severity scores>
}
\end{promptverbatim}
\end{prompttemplate}

\begin{prompttemplate}{Phase 2b --- specificity auditor, system}
\begin{promptverbatim}
You are a privacy specificity auditor. Judge whether the rewritten transcript still reveals participant attributes at a too-specific level. Use the provided dimension definitions and examples. Output valid JSON only.
\end{promptverbatim}
\end{prompttemplate}

\begin{prompttemplate}{Phase 2b --- specificity auditor, user}
\begin{promptverbatim}
Evaluate whether this rewritten transcript is still too specific on the dimensions below.

Mark too_specific=true only when details are specific enough to materially increase identity risk. Mild abstraction loss is acceptable.

Dimensions:
- OCCUPATION (Occupational Domain)
  Too specific example: "teaches cooking classes at a local Italian supermarket chain"
  Just right example: "works in food education and catering"
  Too vague example: "works in a service industry"
- EXPERTISE (Expertise & Experience Level)
  Too specific example: "15-year veteran chef who trained at Le Cordon Bleu"
  Just right example: "experienced professional with deep domain expertise"
  Too vague example: "someone with work experience"
- AI_USAGE (AI Interaction Context)
  Too specific example: "uses Claude to write menus for wedding catering events"
  Just right example: "uses AI for professional writing and content creation in their field"
  Too vague example: "uses AI at work"
- REGION_TYPE (Geographic/Cultural Context)
  Too specific example: "lives in St. John's, Newfoundland and references local festivals"
  Just right example: "based in a coastal Canadian city with strong local food culture"
  Too vague example: "lives somewhere in North America"
- LIFE_STAGE (Life Stage & Demographics)
  Too specific example: "42-year-old married father of two who recently changed careers"
  Just right example: "mid-career working adult with a family"
  Too vague example: "an adult"

=== REWRITTEN TRANSCRIPT ===
{rewritten_text}
=== END REWRITTEN TRANSCRIPT ===

Return JSON with this structure:
{
  "dimensions": {
    "OCCUPATION": {"too_specific": bool, "rationale": "...", "evidence_spans": ["..."]},
    "EXPERTISE": {"too_specific": bool, "rationale": "...", "evidence_spans": ["..."]},
    "AI_USAGE": {"too_specific": bool, "rationale": "...", "evidence_spans": ["..."]},
    "REGION_TYPE": {"too_specific": bool, "rationale": "...", "evidence_spans": ["..."]},
    "LIFE_STAGE": {"too_specific": bool, "rationale": "...", "evidence_spans": ["..."]}
  },
  "too_specific_count": <int>
}
\end{promptverbatim}
\end{prompttemplate}

\subsection{Phase 2c: keeper}
\label{app:prompt:keeper}

\begin{prompttemplate}{Phase 2c --- system}
\begin{promptverbatim}
You are a qualitative research analyst evaluating whether a privacy-rewritten transcript preserves the research-valuable content of the original.

You have access to:
- The ORIGINAL transcript (ground truth)
- The REWRITTEN transcript (privacy-protected version)
- The MASK MAP showing what was replaced

For each of the 8 utility dimensions, assess:
1. What key content existed in the original?
2. Was it preserved, distorted, or lost in the rewrite?
3. If lost, what specifically was lost and how severe is the loss?

Be precise: cite exact spans from both texts to support your assessment.
Output valid JSON only.
\end{promptverbatim}
\end{prompttemplate}

\begin{prompttemplate}{Phase 2c --- user}
\begin{promptverbatim}
=== ORIGINAL TRANSCRIPT ===
{original_text}
=== END ORIGINAL ===

=== REWRITTEN TRANSCRIPT ===
{rewritten_text}
=== END REWRITTEN ===

=== MASK MAP (original -> replaced) ===
{mask_map}

=== UTILITY ATTRIBUTES TO EVALUATE ===
- THEME (Thematic Content): Core topics, recurring themes, and subject domains
- EXPER (Experiential Narratives): Specific events, stories, anecdotes
- AFFECT (Emotional/Affective Expressions): Feelings, attitudes, frustrations
- REASON (Reasoning & Beliefs): Opinions, justifications, decision rationale
- BEHAV (Behavioral Patterns): Workflows, habits, routines
- RELAT (Relational Dynamics): Interactions with others, social roles
- TEMPO (Temporal Structure): Chronology, turning points, development
- EXPERT (Domain Knowledge): Professional/technical insights, vocabulary

For EACH attribute, provide:
- preserved: true/false
- loss_severity: 1-5 (1 = fully preserved, 5 = completely destroyed)
- original_content: brief description
- rewritten_content: brief description
- lost_details: list of specific content items lost or distorted
- recovery_suggestion: how the refiller could recover lost content

Return JSON:
{
  "THEME": {"preserved": bool, "loss_severity": int, ...},
  ... (all 8 attributes)
  "total_loss": <sum of all loss_severity scores>
}
\end{promptverbatim}
\end{prompttemplate}

\subsection{Utility-Benchmark Prompt: Interviewee-profile Fact Generation}
\label{app:profile_fact_prompt}

The Interviewee-profile benchmark uses a separate evaluation pipeline.
After generating and cleaning one attribute summary per profile dimension, that pipeline decomposes each supported summary into non-overlapping atomic facts. The prompt below is the atomic fact-generation prompt used in that second step.

\begin{prompttemplate}{Interviewee-profile attribute summary --- system}
\begin{promptverbatim}
You are a careful qualitative researcher. Given an interview transcript, produce one concise attribute summary for the participant using only evidence from the transcript. When the attribute is supported, prefer a richer and more informative summary rather than a minimal label. Do not use outside knowledge. Return valid JSON only.
\end{promptverbatim}
\end{prompttemplate}

\begin{prompttemplate}{Interviewee-profile attribute summary --- user}
\begin{promptverbatim}
Read the interview transcript and summarize only the participant's {attribute_display_name}.

{attribute_note}

Attribute boundary rules:
{attribute_boundary_rules}

Rules:
- Produce exactly one description for this single attribute.
- Use only evidence from the transcript.
- When supported, make the description longer and more detailed than a bare label.
- Prefer a compact but information-dense noun phrase or one sentence fragment of about 12-35 words.
- Detailed does not mean broad: include only specifics that truly belong to this attribute and exclude content that belongs to the other 7 attributes.
- Preserve uncertainty markers like likely, possible, unclear, or affiliated with when the evidence is partial.
- Do not overclaim or turn weak hints into definite statements.
- If the strongest evidence mainly supports a different attribute, do not use it here; use the unknown fallback instead.
- Do not use generic outputs like `College Degree` when a more detailed evidence-grounded summary is possible; but if the transcript does not directly support the attribute, use the unknown fallback instead of guessing.
- If the attribute is missing, ambiguous, or unsupported, set `description` to: "Unknown {target_attribute_str}; the transcript does not provide enough evidence."
- Set `supported` to false when using the unknown fallback, otherwise true.
- If supported, provide one exact evidence quote from the transcript.
- If unsupported, `evidence_quote` must be an empty string.

Example style for a supported attribute summary:
- `Astrophysicist and numerical modeler of colliding radiative plasma flows (University of Rochester), first author of the cooling-and-instabilities simulation series`

Transcript ID: {transcript_id}

=== TRANSCRIPT ===
{transcript_text}
=== END TRANSCRIPT ===

Return JSON exactly: {"description":"...","supported":true,"evidence_quote":"..."}
\end{promptverbatim}
\end{prompttemplate}

\begin{prompttemplate}{Profile fact generation, system}
\begin{promptverbatim}
You are a careful privacy-analysis assistant. Given one attribute summary description, break it into distinct atomic facts for that attribute only. Use only the provided summary description. Do not use outside knowledge. Return valid JSON only.
\end{promptverbatim}
\end{prompttemplate}

\begin{prompttemplate}{Profile fact generation, user}
\begin{promptverbatim}
Break the following {display_name} summary into non-overlapping atomic facts.
{attribute_note}
Attribute boundary rules:
{boundary_rules}
Rules:
- Use only the summary description.
- Return facts for this attribute only.
- Facts must be distinct, atomic, and non-overlapping.
- If the summary contains mixed information, keep only the parts that actually belong to this attribute and discard the rest.
- If the summary says the attribute is unknown, unsupported, or lacks evidence, return an empty list.
- Do not add outside knowledge.

Summary description: "{summary_description}"

Return JSON exactly: {"facts":["..."]}
\end{promptverbatim}
\end{prompttemplate}

\begin{prompttemplate}{Codebook fact construction --- No Prompt}
\begin{promptverbatim}
No LLM prompt is used for codebook fact construction. The script deterministically reads each human-coded report row, selects rows with a valid code ID plus non-empty original excerpt and codebook note, and stores the human-authored field as the reference code fact together with its code ID, code label, category, code definition, inclusion criteria, exclusion criteria, and source excerpt.
\end{promptverbatim}
\end{prompttemplate}

\begin{prompttemplate}{Codebook fact recoverability --- system}
\begin{promptverbatim}
You are a careful qualitative researcher. Judge whether a specific qualitative code fact can be recovered from a transcript. Use only the transcript text. Do not use outside knowledge. Return valid JSON only.
\end{promptverbatim}
\end{prompttemplate}

\begin{prompttemplate}{Codebook fact recoverability --- user}
\begin{promptverbatim}
Determine whether the qualitative code fact below can be recovered from the transcript for config `{config_label}`.

Transcript ID: {transcript_id}
Category: {category_id} ({category_label})
Code: {code_id} ({code_label})
Code definition: "{code_definition}"
Inclusion criteria: "{inclusion_criteria}"
Exclusion criteria: "{exclusion_criteria}"
Reference excerpt from original transcript: "{excerpt}"
Reference code fact: "{fact}"

Allowed decisions (must match exactly one of these strings):
- Yes, the answer can be recovered from the transcript with clear evidence
- No, the answer cannot be recovered because the transcripts contain ambiguous information and not specific enough to answer the question
- I am not sure

Rules:
- Judge recoverability from the transcript alone.
- The code metadata and reference excerpt clarify the target meaning, but they are not evidence for the current transcript.
- Choose `Yes...` only when the transcript contains clear evidence for the same substantive coded insight.
- Choose `No...` when the transcript is missing, weaker, broader, or not specific enough to support the code fact.
- Choose `I am not sure` only when the transcript hints at the code fact but the evidence quality is genuinely unclear.
- If the decision is Yes, provide one exact evidence quote from the transcript.
- If the decision is No or I am not sure, `evidence_quote` may be empty.

=== TRANSCRIPT ===
{transcript_text}
=== END TRANSCRIPT ===

Return JSON exactly: {"decision":"...","reasoning":"...","evidence_quote":"..."}
\end{promptverbatim}
\end{prompttemplate}

\subsection{Privacy-Benchmark Prompt: Direct-intent Re-identification}
\label{app.direct_intent_prompt}

This privacy benchmark use the following prompt and GPT-5.1 (web-search, reasoning level high), GPT-5.4-mini (web-search, reasoning level high), and Gemini-3-flash-preview (web-search, reasoning level high) to conduct the re-identification. The re-identification results were manually checked on the expert-verified ground truth. The same prompt are used in the adaptive privacy attribute generation. The "matches\_description" of the original text are fed into an LLM as context to report the vulnerability against web-search re-identification attack. The attributes generation prompt is in \ref{AP_prompt}.

\begin{prompttemplate}{Direct-intent re-identification --- system}
\begin{promptverbatim}
I have an interview transcript. Based on the content, try your best to guess who this person might be. Consider all clues including their role, experience, projects they mentioned, opinions, writing style, and any other details that might help identify them. Be thorough and consider how different pieces of information might combine to point to a specific individual.

Rules:
- Be conservative and precise; avoid hallucinations. Prefer real links and include associated details when possible.
- Use web search to validate details whenever helpful.
- Return STRICT JSON matching the schema exactly; do not include any extra commentary.

Output: Return STRICT JSON with the following schema:
{
  "identifier_candidates": [
    {
      "interviewee_candidates": array of strings (if known),
      "identity_label": string,
      "platform_or_source": string | null,
      "link": string,
      "year": integer | null,
      "confidence": "very low" | "low" | "medium" | "high" | "very high",
      "matches_description": string,
      "mismatches_description": string
    }
  ]
}

Guidance:
- The goal is to find the most likely publicly visible item that matches the description; stop early if confidence is very high.
- Make "confidence" reflect how strongly the identified item aligns with the provided description, given evidence.
- In "matches_description", cite specific phrases or details that align with the description.
- In "mismatches_description", call out missing or contradictory details vs. the description.
- If unsure, include candidates with "very low" confidence and explain why.
\end{promptverbatim}
\end{prompttemplate}

\begin{prompttemplate}{Direct-intent re-identification --- user}
\begin{promptverbatim}
Here is the transcript:
{full_transcript}
\end{promptverbatim}
\end{prompttemplate}

\subsection{Adaptive Privacy AURA Prompt: Attribute Generation}
\label{AP_prompt}

\begin{prompttemplate}{Adaptive privacy scope expansion --- user}
\begin{promptverbatim}
Base attributes already present (DO NOT repeat these or close synonyms):
{base_attributes_json}

Transcript ID: {transcript_id}

Original transcript excerpt:
{transcript_excerpt}

Re-identification evidence (top candidates with confidence, matches, and mismatches):
{reid_evidence_json}

Task:
Propose additional privacy attributes beyond the base list that directly capture the specific evidence used to re-identify this transcript.

Priorities:
1) Produce HIGH-LEVEL categories that group related quasi-identifiers.
   Good: RESEARCH_AREA (specific topics/methodologies/subfields)
   Good: TOOL_STACK (software/platform/framework mentions)
   Good: PUBLICATION_SIGNATURE (paper titles/venues/co-author patterns)
   Good: ORG_TYPE (employer or organization category)
   Good: CAREER_STAGE (early-career/mid-career/senior patterns)
   Bad: SPECIFIC_PAPER_TITLE, EXACT_TOOL_VERSION, INDIVIDUAL_PROJECT_NAME
2) Merge overlapping quasi-identifiers into a single broader attribute.
3) Avoid micro-level singleton attributes tied to one exact proper noun.
4) Ensure each attribute is actionable for a masker (generalizable, not just removable).
5) Return at most {max_attributes} attributes.

Return STRICT JSON with this schema:
{
  "attributes": [
    {
      "key": "UPPERCASE_SHORT_KEY",
      "display_name": "Human-friendly name",
      "target_str": "what this attribute targets in text",
      "options": ["optional", "categorical", "values"],
      "note": "optional note"
    }
  ]
}

Rules:
- All returned attributes must be different from the base 8.
- Avoid near-duplicates of each other.
- Keep key length <= 16 chars.
- If no strong additions exist, return an empty list.
\end{promptverbatim}
\end{prompttemplate}